\documentclass[a4paper,11pt]{article}
\pdfoutput=1 % if your are submitting a pdflatex (i.e. if you have
\usepackage{jcappub} % for details on the use of the package, please
\usepackage[T1]{fontenc} % if needed
\usepackage{hyperref}
\usepackage{url}
\usepackage{multirow}
\usepackage{graphicx}
\usepackage[dvipsnames]{xcolor}
\newcommand{\LV}[1]{\textcolor{orange}{[{\bf LV}: #1]}}

\title{\boldmath  BE-HaPPY: Bias Emulator for Halo Power Spectrum Including Massive Neutrinos}

\definecolor{darkred}{RGB}{175,0,0}

\author[a,1]{David Valcin,}\note{Corresponding author.}
\author[b]{Francisco Villaescusa-Navarro,}
\author[c,a]{Licia Verde}
\author[d,a]{Alvise Raccanelli}

\affiliation[a]{ICC, University of Barcelona, IEEC-UB, Marti i Franques, 1, E08028 Barcelona, Spain}
\affiliation[b]{ Center for Computational Astrophysics, Flatiron Institute, 162 5th Avenue, 10010, New York, NY, USA}
\affiliation[c]{ICREA, Pg. Lluis Companys 23, 08010 Barcelona, Spain}
\affiliation[d]{Theoretical Physics Department, CERN, 1 Esplanade des Particules, CH-1211 Geneva 23, Switzerland}

\emailAdd{d.valcin@icc.ub.edu}
\emailAdd{fvillaescusa@flatironinstitute.org}
\emailAdd{liciaverde@icc.ub.edu}
\emailAdd{alvise.raccanelli@cern.ch}

\abstract{We study the clustering properties of dark matter haloes in real- and redshift-space in cosmologies with massless and massive neutrinos through a large set of state-of-the-art N-body simulations. We provide quick and easy-to-use prescriptions for the halo bias on linear and mildly non-linear scales,  both in real and redshift-space,  which are valid also for massive neutrinos cosmologies. Finally we present a halo bias emulator,~\textbf{BE-HaPPY}, calibrated on the N-body simulations, which is fast enough to be used in  the standard Markov Chain Monte Carlo approach to cosmological inference. 
For a fiducial standard $\Lambda$CDM cosmology \textbf{BE-HaPPY} reproduces the simulation inputs with percent or sub-percent accuracy  for  the halo mass cuts  it is calibrated on ($M>\{5\times 10^11, 10^{12}, 3\times 10^{12}, 10^{13}\} h^{-1} M_{\odot}$)  on the  scales of interest (linear and well into the mildly non-linear regime). The approach presented here represents a well defined route to  meeting  the halo-bias  accuracy requirements  for the analysis of next-generation large--scale structure surveys.  The software  \textbf{BE-HaPPY} can run both in emulator mode and in calibration mode, on user-supplied simulations outputs, and is made publicly available.  }

\subheader{\footnotesize
        \vspace*{-3.7em}
        \begin{flushright}
                CERN-TH-2018-223
        \end{flushright}
}
\begin{document}
\maketitle
\flushbottom

%%%%%%%%%%%%%%%%%%%%%%%%%%%%%%%%%%%%%%%%%%%%%%%%%
%%%%%%%%%%%%%%%%%%%%%%%%%%%%%%%%%%%%%%%%%%%%%%%%%
\section{Introduction}
To fully take advantage of next generation surveys such as Euclid\footnote{http://sci.esa.int/euclid/}, DESI\footnote{http://desi.lbl.gov/etc}, WFIRST\footnote{https://wfirst.gsfc.nasa.gov/}, SKA\footnote{https://www.skatelescope.org/} EMU\footnote{https://www.atnf.csiro.au/people/Ray.Norris/emu/index.html}, PSF\footnote{https://pfs.ipmu.jp},  and LSST\footnote{https://www.lsst.org/}  we must improve our modelling of clustering of the tracers of the dark matter density field. The amplitude and scale dependence of the matter power spectrum carry important cosmological informations about e.g., the primordial Universe or the absolute neutrino mass scale, highly complementary to that provided by cosmic microwave background
observations. Galaxy or halo bias, which is the relation between these tracers and the underlying matter field, is one of the main source of uncertainty preventing us from achieving this goal.
Since galaxies are hosted in dark matter halos, the first step is to model correctly the bias of the halo field or halo bias. Hereafter when we refer to bias we mean the halo bias. Accurate modelling of the halo bias is a necessary (although not sufficient) step to achieve accurate modelling of the  observable dark matter tracers. The (halo) bias is usually
approximated by a constant on linear scales and then marginalized over. However the approximation of scale independence may be insufficient, even on linear scales. This is all the more true in a cosmological model with massive neutrinos. Indeed, because of their thermal
velocities, neutrinos act as relativistic species during the growth of  cosmological perturbations and therefore can escape region of higher density fluctuations. This phenomenon, known as the ``neutrino streaming" effect, results in suppression of power at small scales. Massive neutrinos also have
an additional effect on the growth of structures. As tiny as their mass could be, neutrinos modify the shape of the power spectrum and thus the halo bias. Neutrinos are one of the most mysterious fundamental particles of nature. The value of their masses remains a mystery  today. Constraining their masses is among the goals of upcoming surveys. In order to achieve this, accurate theoretical predictions are needed. The purpose of this work is to investigate in detail the shape and amplitude of the halo bias, as a proxy and a  preliminary step for galaxy bias, in cosmologies with massive neutrinos into the mildly non-linear\footnote{Here mildly non-linear scales refers to scales where non linear effects arise but low order perturbation-theory approximations are still valid.  } and non-linear regime, and offer a fast way to model it.

While not an issue  for present-day surveys, Raccanelli et al. \cite{1} (see also  Vagnozzi et al. \cite{2}), showed that an inaccurate model for the  bias in cosmologies with massive neutrinos will induce a systematic and statistically significant  shift in the  inferred cosmological parameters for forthcoming  surveys.

A solution proposed by e.g., \cite{1,3,4,5,6} to
account for this massive neutrinos effect is to use the power spectrum of the cold dark matter plus baryons, $P_{\rm cc}$, instead of  that of the total matter, $P_{\rm mm}$, as the relevant theoretical input. It is therefore  $P_{\rm cc}$  the quantity to be modelled and thus the one to be used in the definition  the tracers bias. On large-scales, in cosmologies with massive
neutrinos, the halo bias defined in this way become effectively scale-independent and on smaller scales, its scale-dependence, has been found to be neutrino-mass independent \cite{4,5,6} (at least to current precision); a small scale dependence even on linear scales is expected \cite{7}, but it  does not affect the results presented here.

In this work, we use a large set of state-of-the-art N-body simulations with massive and massless neutrinos to study and model the effects induced by massive neutrinos on halo bias.  We establish a  simple link between the halo bias in models with massive and massless neutrinos. The results of this investigation are summarised in a  software package which computes halo bias including  its scale dependence, also  in the presence of massive neutrinos,  \textbf{BE-HaPPy}: Bias Emulator for Halo Power spectrum in Python. \textbf{BE-HaPPy} provides a  bias emulator, fast enough to be used as a plug-in for standard Markov Chain Monte Carlo (MCMC) cosmological analyses, which is accurate, easy to implement and signifies only a small additional computational cost.  With  \textbf{BE-HaPPy}  a standard Boltzmann-MCMC  can  quickly compute also  the halo power spectrum into the mildly non-linear regime. While  strictly we have calibrated the bias emulator for a fixed set of cosmological parameters, those of a standard concordance LCDM model, we will argue that current data already constrain cosmological parameters enough that the \textbf{BE-HaPPy} approach  can be used  beyond the specific cosmology used here. Nevertheless  \textbf{BE-HaPPy}   can also  be run in calibration mode with a user-supplied set of power spectra for arbitrary cosmologies.

 Calibration on simulations is not the only approach that has been proposed in the literature. Recently, Mu\~{n}oz and Dvorkin \cite{3} also studied the impact of massive neutrinos in the galaxy bias and,
as \cite{1}, reached the conclusion that their effect should be included in any future survey analysis.
They developed a code \textbf{RelicFast}, \cite{3} which computes the large, linear scales Lagrangian and Eulerian biases in the
presence of relics that are non-relativistic today (see \cite{6,7} for some background on this topic).
\textbf{RelicFast} and \textbf{BE-HaPPy} offer two complementary codes to compute the halo bias in the presence of massive neutrinos. 
\textbf{RelicFast} offers quasi-analytical approach to compute the large-scales scale-dependence of the linear bias through spherical collapse and peak-background split, where \textbf{BE-HaPPy}
uses fitting and interpolating functions calibrated on N-body simulations on linear-to-mildly non-linear scales. Simulations are less
versatile (only a finite set of cosmologies can be explored) but remain one of the best method
to obtain the bias especially in the (mildly)non-linear regime. The analytical approach offers valuable physics insights but is valid only on fully linear scales; hence the two approaches are highly complementary.
This paper is structured as follows. After an introduction to notation, definitions and set up in Sec.~\ref{sec:Definitions}, we briefly present the tools we used to study and model the halo bias. In Sec.~\ref{sec:BE} we introduce the methodology and the choices made towards the development of the emulator, which is designed for both cosmologies with massive and massless neutrinos. Our emulator works both in real- and redshift-space. We discuss in detail the extension of our emulator in redshift-space in Sec.~\ref{sec:RSE}. In Sec.~ \ref{sec:BEH} we summarize the main properties and features of our emulator and conclude in Sec.~\ref{sec:conclusion}.

%%%%%%%%%%%%%%%%%%%%%%%%%%%%%%%%%%%%%%%%%%%%%%%%%
%%%%%%%%%%%%%%%%%%%%%%%%%%%%%%%%%%%%%%%%%%%%%%%%%
\section{Definitions, set up and methodology}
\label{sec:Definitions}
%\label{subsec:definitions}
The key idea we build upon is that, in presence of massive neutrinos, halo bias should not be defined with respect to total matter $P_{\rm mm}(k)$, but with respect to the  cold dark matter (CDM)+baryons field, $P_{\rm cc}(k)$:
\begin{equation}
b_{\rm mm}(k) = \sqrt{\frac{P_{\rm hh}(k)}{P_{\rm mm}(k)}} \;\Rightarrow \;b_{\rm cc}(k) = \sqrt{\frac{P_{\rm hh}(k)}{P_{\rm cc}(k)}}\,.
\label{bcc}
\end{equation}
The reason behind  this idea is that neutrinos barely cluster on small scales \citep{15}, so both the abundance and clustering of haloes and galaxies will be characterized by the CDM+baryon density field instead of the total matter field \citep{4,5,1}. We note however that it is expected that the scale-dependent growth rate produced by neutrinos will induce a small linear scale-dependent bias \citep{6,7}. We have neglected this small effect because here we are interested in studying the theoretical templates needed to describe halo clustering on mildly to fully non-linear scales. This effect can be  included a posteriori and on larger scales,  since it affects  $k\lesssim 10^{-2}$ h/Mpc (see \cite{7, chiangPaco} for details).

At linear order, the two halo bias definitions can simply be related through the linear transfer functions
\begin{equation}
b_{\rm mm}(k) = \frac{T_{\rm cc}(k)}{T_{\rm mm}(k)}\:b_{\rm cc}(k) 
\label{bmm1}	
\end{equation}
where
\begin{equation}
T_{\rm cc}(k) = \frac{\Omega_{\rm c}T_{\rm cc}(k) + \Omega_{\rm bb}T_{\rm b}(k)}{\Omega_{\rm c} + \Omega_{\rm b}}\,,
\end{equation}
and the subscripts ${\rm c}$, ${\rm b}$ and ${\rm m}$ stand for CDM, baryons and total matter (i.e. CDM plus baryons plus neutrinos) respectively. $\Omega_i$ represents the energy fraction of each component $i$ at $z=0$. We note that the total matter power spectrum and the different transfer functions can be easily obtained from Boltzmann solvers such as CLASS and CAMB \citep{10,11}. Raccanelli et al. \citep{1} showed that the validity of the above equation extends well into the (mildly) non-linear regime. 

In this paper we will be working under one important assumption: neutrinos only affect the overall amplitude of the bias ($b_{\rm cc}$), not its scale-dependence
\begin{equation}
b_{\rm cc}(k, M_\nu) \simeq \alpha \, b_{\rm cc}(k,M_\nu=0)=\frac{b_{\rm cc}^{\rm LS}(M_{\nu})}{b_{\rm cc}^{\rm LS}(M_{\nu}=0)} b_{\rm cc}(k,M_\nu=0)~,
\label{eq:rescaling}
\end{equation}
 where we have followed the notation of Ref.~\cite{1} and $b_{\rm cc}^{\rm LS}$ denotes the large-scale bias for CDM+baryons;  $b_{\rm cc}^{\rm LS}$ is computed on linear scales where $b_{\rm cc}$ becomes scale-independent ($b_{\rm cc}^{\rm LS}$ can be interpreted as the limit\footnote{If the small effect --evident on scales larger than $k\simeq 10^{-2}$h/Mpc-- of a  scale dependence of the linear bias of Ref.~\cite{6}  is to be  included in the modelling,  then $b_{\rm cc}^{\rm LS}$ should be computed on  large linear scales where  the bias "plateau"  is \cite{6, 7, chiangPaco}.}   of  $b_{\rm cc}(k)$ for $k\longrightarrow 0$).

The above equation relates the halo bias between two models that have the same values for the parameters $h$, $n_s$, $\Omega_{\rm m}$, $\Omega_{\rm b}$ and $A_s$, but different values of $\Omega_{\rm c}$ and neutrino mass,  where $\Omega_{\rm c} = \Omega_{\rm m} - \Omega_{\rm b} - \Omega_\nu$. 
Conveniently, the scale dependence of $b_{\rm cc}$ can be computed for massless neutrino cosmology. This has two immediately obvious advantages: it can be calibrated on massless neutrino simulations, which are easier to run, and it can be modelled, for example, by resorting to a perturbation theory description of the power spectrum, which validity has been studied extensively for massless neutrinos cosmologies and which can be computed given a set of  cosmological parameters.

Equation \ref{eq:rescaling} is an approximation that is expected to break down if the neutrino masses are large and/or if the halo bias is high, see Ref.~\cite{1}. Below we will test the performance and exploit the potential of the above equation.

While we  will be focusing our attention on modelling $b_{\rm cc}(k)$ in cosmologies with massive and massless neutrinos, if, in models with massive neutrinos, the desired quantity is the halo bias with respect to the total matter density field, it can easily be obtained from Eqs. \ref{bmm1} and \ref{eq:rescaling} as
\begin{equation}
b_{\rm mm}(k,M_{\nu} ) = \frac{T_{\rm cc}(k)}{T_{\rm mm}(k)}\frac{b_{\rm cc}^{\rm LS}(M_{\nu})}{b_{\rm cc}^{\rm LS}(M_{\nu}=0)}b_{\rm cc}(k, M_{\nu} = 0)~.
\label{bmm2}	
\end{equation}
 
Below we will present two approaches to model  $b_{\rm cc}$: one phenomenological polynomial model (as in \cite{1}) and one perturbation theory-based; each will be calibrated on simulations.

\subsection{N-body simulations}
\label{sec:sims}
The N-body simulations analyzed in this paper belong to the HADES suite (initially presented in \citep{41} but extended since). They were run using the TreePM+SPH code \textsc{Gadget-III}, (see \citep{12} for a description of \textsc{Gadget-II}). The simulations follow the evolution of $1600^3$ CDM and $1600^3$ neutrino particles in a box of size 1000 comoving $h^{-1}$Mpc. The gravitational softening of both CDM and neutrinos is set to 15 $h^{-1}$kpc. All simulations share the value of the following cosmological parameters, that are in excellent agreement with the latest constraints from Planck \cite{13}: $\Omega_{\rm m}=\Omega_{\rm c}+\Omega_{\rm b}+\Omega_\nu=0.3175$, $\Omega_{\rm b}=0.049$, $\Omega_\Lambda=0.6825$, $\Omega_{\rm k}=0$, $h=0.6711$, $n_s=0.9624$ and $A_s=2.13\times10^{-9}$. In models with massive neutrinos we set $\Omega_\nu h^2=M_\nu/93.14$ eV, where $M_\nu=\sum_i m_{\nu_i}$. We assume three degenerate neutrino masses in our simulations, as neutrino mass hierarchy is not relevant to our approach. 

We use the classical particle-based method \citep{14,15} to simulate the evolution of massive neutrinos in the fully non-linear regime. The initial conditions were generated at $z=99$ through the method illustrated in \cite{16}, i.e., by rescaling the $z=0$ power spectrum and transfer functions while accounting for the scale-dependent growth factor and growth rate present in cosmologies with massive neutrinos. We have run simulations for two different models. A model with massless neutrinos and a model with $M_\nu=0.15$ eV. For each model, we have run 10 paired fixed simulations\footnote{Note that each pair of fixed simulations consists of two simulations} \citep{17, 18}. As shown in \cite{18}, this set up improves the statistics of all clustering measurements considered in this work. While we do not expect improvements for the halo bias, a significant reduction on the sample variance of quantities such as the matter or halo power spectrum can be achieved through this setup (see discussion in Ref.~\cite{17}). 

For each simulation we have saved snapshots at redshifts 0, 0.5, 1 and 2. Dark matter haloes are identified through the Friends-of-Friends algorithm \citep{19} with a value of the linking length parameter equal to $b_{l}=0.2$. Our halo catalogues consists of all haloes with masses above $5\times10^{11}~h^{-1}M_\odot$. Smaller halos would not have a sufficient number of particle to provide a sufficiently converged halo power spectrum. In
reality to study the halo-halo correlation properties at mildly non-linear scales it is customary to consider a minimum number of particles per halo around few tens because at this level the halo correlation function at large scales is expected to be sufficiently converged. We are
consistent with this convention.

\subsection{Halo mass bins and $k_{\rm max}$}
Since the halo bias depends on halo mass, we consider four different halo mass bins. Instead of focusing on narrow mass bins, where our statistics will be limited, we consider all haloes above a certain mass. We work with haloes with masses above $5\times10^{11}~h^{-1}M_\odot$ (M1), $1\times10^{12}~h^{-1}M_\odot$ (M2), $3\times10^{12}~h^{-1}M_\odot$ (M3) and $1\times10^{13}~h^{-1}M_\odot$ (M4). The different mass bins are also shown in Table \ref{mr}. We do not consider mass bins with a higher  mass cut given their very low number density in both simulations and data. 

Another important parameter in our analysis is the minimum scale --maximum wavenumber-- used, $k_{\rm max}$. The amount of information that can be extracted from galaxy surveys depends critically on $k_{\rm max}$,  however modelling becomes increasingly complicated and less accurate with increasing $k$. We explore the performance of our approach as a function of $k_{\rm max}$.
 In particular, following \cite{1}, we  also consider the  three different cases (I, II and III) for $k_{\rm max}$. For case I the maximum $k$ increases in redshift so that the r.m.s of the density fluctuations is constant in redshift and has the same value as the one for $k_{\rm max}$ = 0.16 h/Mpc at z = 0. Case II is more conservative, having $k_{\rm max}$ = 0.12 h/Mpc at z = 0; $k_{\rm max}$ initially grows in redshift to keep $\Delta^2$ ($k_{\rm max}$) constant but then it saturates at $k_{\rm max}$ = 0.2 h/Mpc. Case III is simpler and  conservative, as it keeps $k_{\rm max}$ = 0.15 h/Mpc, constant in redshift. Table \ref{mr2} summarizes the different cases.

\begin{table}
\centering
\begin{tabular}{|c|c|c|c|c|}
\hline
bin name & M1 & M2 & M3 & M4 \\
\hline
mass range ($ h^{-1} M_{\odot} $) & \textgreater $5\times10^{11}$ & \textgreater $1\times10^{12}$ & \textgreater $3\times10^{12}$ &  \textgreater $1\times10^{13}$ \\
\hline
\end{tabular}
\caption{This table shows the different mass bins we have considered in our analysis.}
\label{mr}
\end{table}

\begin{table}
\centering
\begin{tabular}{|c|c|c|c|}
\hline
case  & I & II & III \\
\hline
\multirow{ 2}{*}{$k_{\rm max}$} & \multirow{ 2}{*}{$\bigtriangleup^2(k_{\rm max},z)=\bigtriangleup^2(0.16\, h {\rm Mpc^{-1}}, z)$}  & $\bigtriangleup^2(k_{\rm max},z)=\bigtriangleup^2(0.12\,h {\rm Mpc^{-1}}, z)$ & 0.15 h/Mpc \\
& & \& $k_{\rm max}$ < 0.2 h/Mpc & at all $z$ \\
\hline
\end{tabular}
\caption{This table shows different criteria used to set $k_{\rm max}$.}
\label{mr2}
\end{table}

\subsection{Shot-noise correction}
\label{shot-noise}

The discreteness of haloes affects their measured clustering. To model the cosmological clustering of these tracers,  we need to separate halo discreteness effects from the cosmic signal in our measurements. 

\begin{figure*}
\begin{center}
\includegraphics[scale=0.47]{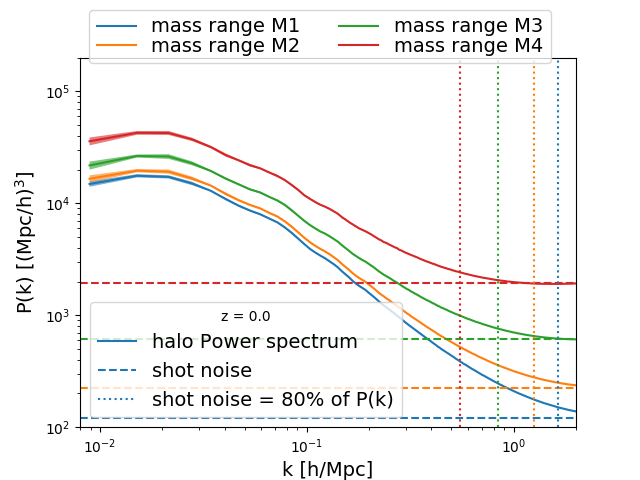}
\includegraphics[scale=0.47]{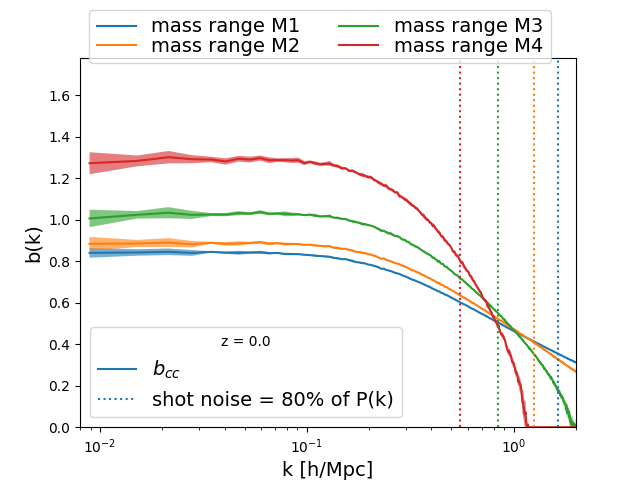}
\caption{\textit{Left}: Halo power spectrum for the massless neutrino model for different mass ranges at $z=0$. We show the mean and the standard deviation of the 10 pairs of different realizations. The expected Poissonian shot-noise contribution, $1/n$ is shown with horizontal dashed  lines for the different mass bins. \textit{Right}: Halo bias after subtracting the haloes shot-noise from their auto-power spectrum. On small scales the shot-noise becomes sub-Poissonian for the most massive halos. In this work we restrict our analysis to scales where the amplitude of the shot-noise is smaller than 80\% of the total halo power spectrum (i.e., $nP>0.25$, vertical dotted lines).}
\label{sn}
\end{center}
\end{figure*}

A simple way to do this is by subtracting a Poisson shot-noise $1/n$, where $n$ is the tracer mean number density, from the measured halo auto-power spectrum. In  the left panel of  Fig.~\ref{sn}  we show the halo power spectrum for the model with massless neutrinos at $z=0$ for different mass bins. In the same panel we display with dashed lines the expected amplitude of the shot-noise. As can be seen, on small scales, the halo power spectrum is dominated by shot-noise, whose amplitude matches well with the expected $1/n$ value.

In the right panel of Fig. \ref{sn} we plot the halo bias; the amplitude of the halo auto-power spectrum is corrected for shot-noise as explained above. The shot-noise contribution to the halo power spectrum can become sub-Poissonian for the most massive haloes \citep{20,21,22}. This effect can be explained by the fact that the more massive haloes occupy a larger volume, implying a halo exclusion mechanism that leads to a sub-Poissonian shot-noise. Under these circumstances, the simple Poissonian shot-noise removal will result in unphysical, negative values for the halo power spectrum.  

In what follows we still subtract a Poisson contribution to correct for the shot-noise\footnote{As it will be clearer later when performing parameters fit the shot noise amplitude will be corrected by  a nuisance parameter to be marginalised over, see appendix~\ref{sec:tuning}.}, but to make sure that sub-Poisson effects do not severely impact our results,  we restrict our analysis to scales where the amplitude of the shot-noise is less than 80\% of the total halo power spectrum. In terms of the widely used $nP$ quantity, where $n$ is the average tracers number density  and $P$  the shot-noise subtracted power spectrum,  we impose $nP>0.25$. The vertical dotted lines in Fig.~\ref{sn} indicate the corresponding scale.

This criteria sets  a limit on the  smallest scale (largest wavenumber $k_{\rm lim}$) we can consider,  which is well into the non-linear regime in all cases: e.g., $k_{\rm lim} \sim 0.55 \, h$/Mpc for  the most stringent case of mass bin M4 at $z =0$. As it will be clear below, the scales of interest for our emulator satisfy $k<k_{\rm lim}$.

\subsection{Perturbation theory}

For a given cosmological model our emulator also computes and provides  the  perturbation-theory prediction of the halo power spectrum.  For this  we use FAST-PT \citep{8,9}.

FAST-PT offers a computationally efficient way to compute the power spectrum (both of dark matter in real and redshift-space and of biased tracers) through perturbation theory and  includes bias up to second order. In our analysis we will also consider third order bias, so we modified FAST-PT to achieve this. The use of a perturbation theory approach such as FAST-PT  ensures that \textbf{BE-HaPPy} can be used beyond the specific cosmology adopted here.

We note that care must be taken when comparing predictions from FAST-PT versus simulation outputs.  The FAST-PT input power spectrum must be precisely  sampled;  uneven sampling due to a finite number of significant digits will appear as numerical noise \cite{8}. We apply the same $k$-binning to both the output of FAST-PT and the simulations. This provides a fair comparison among the two results and avoid artificial differences due to binning, that can be important on large-scales. 

\subsection{Fitting procedure}

We calibrate out theoretical model by fitting the model parameters to the outputs  of the N-body simulations. For each halo mass range and redshift   the simulations provide the halo  and the CDM+baryons power spectra; we compute the halo bias as
\begin{equation}
b_{\rm cc}(k)=\left[\frac{(P_{\rm hh}(k)-P_{\rm SN})}{P_{\rm cc}(k)}\right]^{1/2}
\label{eq:biassim}
\end{equation}
and estimate its errors from the dispersion of the 10 realizations of each cosmology. We then fit our results using any of the two bias models we consider: a phenomenological polynomial model and perturbation theory. The best-fit (i.e., the multi-dimensional maximum of the posterior) and error bars (actually full posterior distribution) of the theoretical model parameters are found by using a Markov Chain Monte Carlo (MCMC) method.  The  procedure is detailed in Appendix ~\ref{sec:tuning}. 

While to reduce the impact of shot noise   it is customary to define the bias as the ratio between the  halo-matter cross power spectrum and the matter auto power spectrum, here we stick to Eq.~\ref{eq:biassim}. This is motivated by the fact that  beyond a simple linear bias model the two bias definitions may not  coincide.  We argue here that the bias obtained from Eq.~\ref{eq:biassim} is closer to the quantity that will be useful to interpret  clustering observations. In doing so we pay the price  of a higher shot noise.

Due to the limited number of simulations we have access to, our fits do not account for the correlation between different k-bins, i.e., our likelihood only accounts for the diagonal part of covariance matrix. Therefore, the absolute values of the $\chi^2$ should be taken as a mere guide of the quality of the model.

%%%%%%%%%%%%%%%%%%%%%%%%%%%%%%%%%%%%%%%%%%%%%%%%%
%%%%%%%%%%%%%%%%%%%%%%%%%%%%%%%%%%%%%%%%%%%%%%%%%
\section{Halo clustering in configuration space}
\label{sec:BE}

We begin by  studying  in detail the clustering of haloes in real space. We compare and calibrate with  massless neutrino simulations the  two bias models adopted and  then we quantify the accuracy of  our rescaling Eq.~\ref{eq:rescaling} to obtain $b_{\rm cc}(k,M_\nu)$ for the massive neutrinos case from   $b_{\rm cc}(k,M_\nu=0)$.

\subsection{Halo bias model I: polynomial}
\label{sec:scale-dependent-bias}

 It is well known that  the linear, scale independent bias  approximation is  accurate only on very  large-scales \cite{32,33,34,35}. On smaller scales, the bias becomes scale-dependent. Following \cite{1}, we use a simple phenomenological model and parameterize the halo bias as: 
\begin{equation}
b_{\rm cc}(k,z) = b_{1}(z) + b_{2}(z) k^{2} + b_{3}(z) k^{3} + b_{4}(z) k^{4}\,,
\label{plodd}
\end{equation}
where the coefficients $b_1$, $b_2$, $b_3$ and $b_4$ are free-parameters whose values depend on redshift, halo mass, $M_{\nu}$ and cosmology. 
\begin{figure*}
\begin{center}
\includegraphics[scale=0.487]{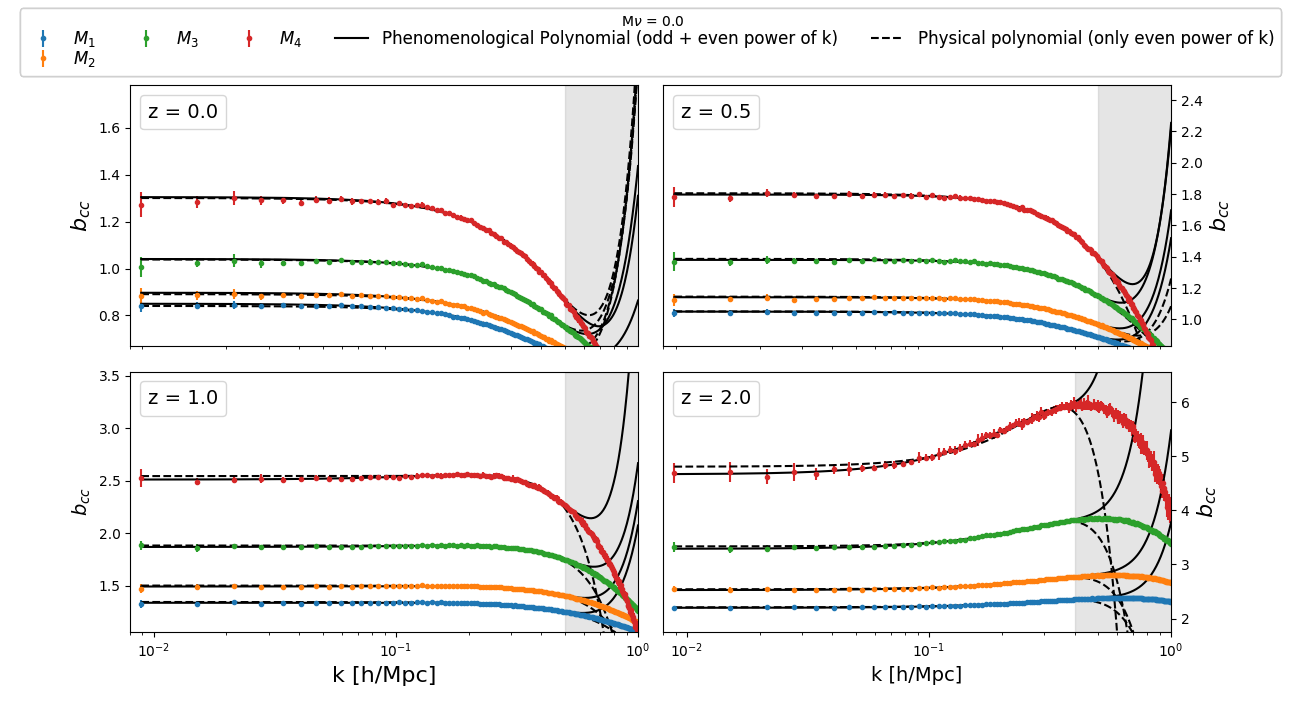}\\
\includegraphics[scale=0.487]{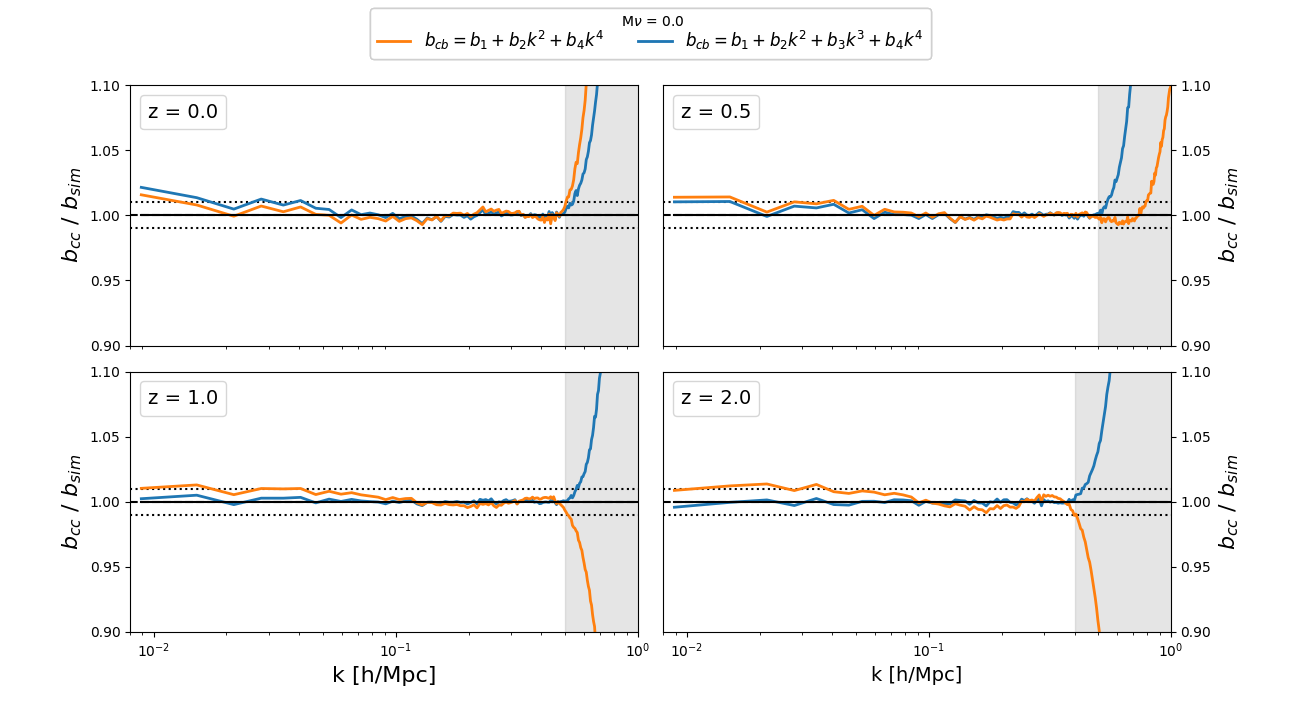}
\caption{The top four panels show the halo bias for the four different mass bins (color lines) at four different redshifts (different panels) for the model with massless neutrinos. The fits for the phenomenological models of Eqs. \ref{plodd} (black solid lines) and \ref{pleven} (black dashed lines) only  include  $k< k_{\rm lim}$ (excluded scales shown as a grey region). The bottom four panels show the ratio between the halo bias obtained from the simulations and the fit. For clarity we only show the average ratio of the four different mass ranges. (Individual mass bins are shown in  appendix \ref{eachmass}, in the fitting range were is not much difference hence justifying showing the mean behaviour). Both models reproduce the halo bias within $\simeq1\%$ in the relevant $k$ range at all redshifts for all mass bins. }
\label{pl1}
\end{center}
\end{figure*}
Eq. \ref{plodd} is however unphysical, as isotropy constraints require the bias to have even powers of $k$ \cite{78}. Nevertheless, we expect that the inclusion of the $k^3$ term improves the quality of the fit. We also use a more physically motivated model with only even powers of $k$:
\begin{equation}
b_{\rm cc}(k,z) = b_{1}(z) + b_{2}(z) k^{2} + b_{4}(z)k^{4}\,.
\label{pleven}
\end{equation}
In both models, the value of the linear (large scale) bias is simply given by $b_{\rm cc}^{\rm LS}(z)=b_1(z)$. 
We fit the halo bias from the  massless neutrinos simulations with the above two models at different redshifts and for the different mass bins. Because of  $k_{\rm lim}$ considerations (see section \ref{shot-noise})  we set $k_{\rm max} =  \{0.55, 0.54, 0.53, 0.42\}$ h/Mpc at $z = \{0, 0.5, 1, 2\}$, respectively.  We show our results in Fig. \ref{pl1}. 

Both approaches yield a very good fit (under $\sim1\%$ until $k_{\rm max}$);  the presence of the extra parameter, $b_3$, slightly improves the quality of the fit on large scales. The best-fit values of the coefficients for all mass bins and redshifts are reported in Appendix \ref{sec:tabelsbiascoeffs1}.
Because of its slightly better fit at the largest scales,  and for direct comparison with Ref.~\cite{1}, unless otherwise stated in what follows our reference ``polynomial" bias model is that of Eq.~\ref{plodd}. The values of the coefficients for the odd-powers polynomial model are provided by \textbf{Be-HaPPy}. This polynomial bias model as calibrated here might not perform as well for different cosmologies. For this reason we introduce below a more flexible bias model that can easily account for varying cosmological parameters.

\subsection{Halo bias model II: perturbation theory}
\label{sec:PTbias}

A more physically motivated model is the perturbation theory-based, non-linear bias expansion \cite{36,37,38,39}.  This approach  has the advantage that the dependence on cosmology is naturally  included. 
 Saito et al. \cite{39} showed that a good model to describe the (shot-noise subtracted) halo power spectrum in N-body simulations in the mildly non-linear regime can be obtained by including up to third-order nonlocal bias terms:
\begin{equation}
\begin{split}
%P_{g,\delta\delta} &= b_{1}^{2}P_{\delta\delta}(k) + 2b_{2}b_{1}P_{b2,\delta}(k)
%\\ 
%&+ 2b_{s2}b_{1}P_{bs2,\delta}(k) + 2b_{3nl}b_{1}\sigma_{3}^{2}P_{\rm lin}(k) 
%\\
%&+ b_{2}^{2}P_{b22}(k) + 2b_{2}b_{s2}P_{b2s2}(k) + b_{s2}^{2}P_{bs22}(k) + P_{\rm SN}
P_{\rm hh}(k) &= b_{1}^{2}P_{\rm cc}(k) + 2b_{2}b_{1}P_{b2,{\rm cc}}(k)
\\ 
&+ 2b_{s2}b_{1}P_{bs2,{\rm cc}}(k) + 2b_{3nl}b_{1}\sigma_{3}^{2}P_{\rm cc}^{\rm lin}(k) 
\\
&+ b_{2}^{2}P_{b22}(k) + 2b_{2}b_{s2}P_{b2s2}(k) + b_{s2}^{2}P_{bs22}(k) %+ %P_{\rm SN}
\end{split}
\label{bpt1}
\end{equation}
where $P_{\rm cc}(k)$ is the non-linear CDM+baryons power spectrum, $P_{\rm cc}^{\rm lin}$ is the linear CDM+baryons power spectrum, $b_{1}$ is the linear bias, $b_{2}$ 2nd-order local bias, $b_{s2}$ 2nd-order non-local bias and $b_{3nl}$ 3rd-order non-local bias\footnote{This term encompasses  various non-local third-order terms. Since it results in a   k-dependent factor that  multiplies  the linear power spectrum,  its contribution  become relevant at ``large'' scales \cite{40} and is therefore considered here. See also Appendix~\ref{appendix:b3}.}. All other terms represent n-loop power spectra (always for CDM+baryons) whose exact expressions can be found in the Appendix \ref{sec:ptterms} or in \cite{37}. The second-order bias expansion consists of all the terms involving the first and second order coefficients $b_{1}, b_{2}$ and $b_{s2}$, while the third order expansion also includes the $b_{3nl}$ term whose explicit expression is reported in Appendix \ref{appendix:b3}. Since FAST-PT does not incorporate this term, we have modified it to account for it.

If  the bias is assumed to be local in Lagrangian space, then the Eulerian bias
is non-local but the  values of  $b_{s2}$ and $b_{3nl}$  are related to  $b_1$: $b_{s2} =-4/7(b_1-1)$ and $b_{3nl} = 32/315(b_1-1)$ \citep{36,37,38,39,40}. Without this constraint, with both $b_{s2}$ and $b_{3nl}$ as free parameters, one accounts for a more general case of a non-local Eulerian bias model. In this work we keep $b_{1}$ and $b_{2}$ as free parameters, and consider two possibilities for $b_{s2}$ and $b_{3nl}$: 1) set them to $-4/7(b_1-1)$ and $\frac{32}{315}(b_{1} - 1)$ (local bias in Lagrangian space), respectively, and 2) leave them as free parameters.

\begin{figure*}
\begin{center}
\includegraphics[scale=0.45]{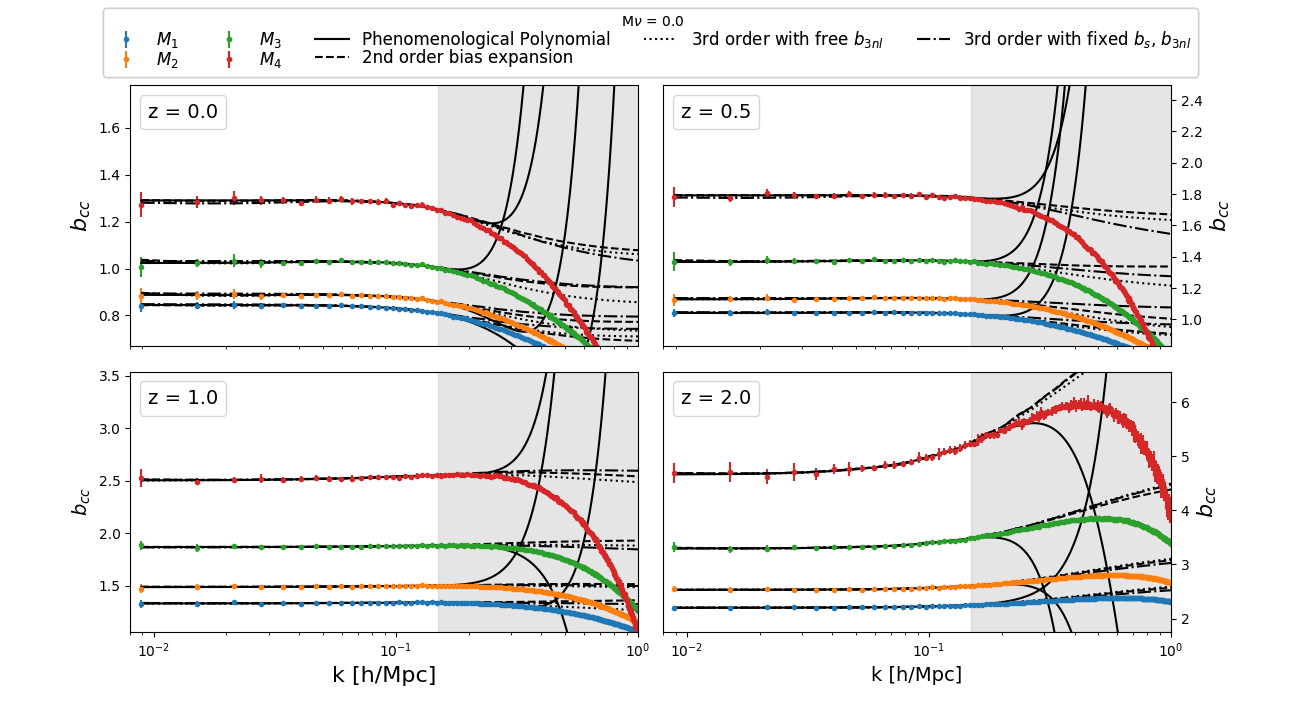}
\includegraphics[scale=0.45]{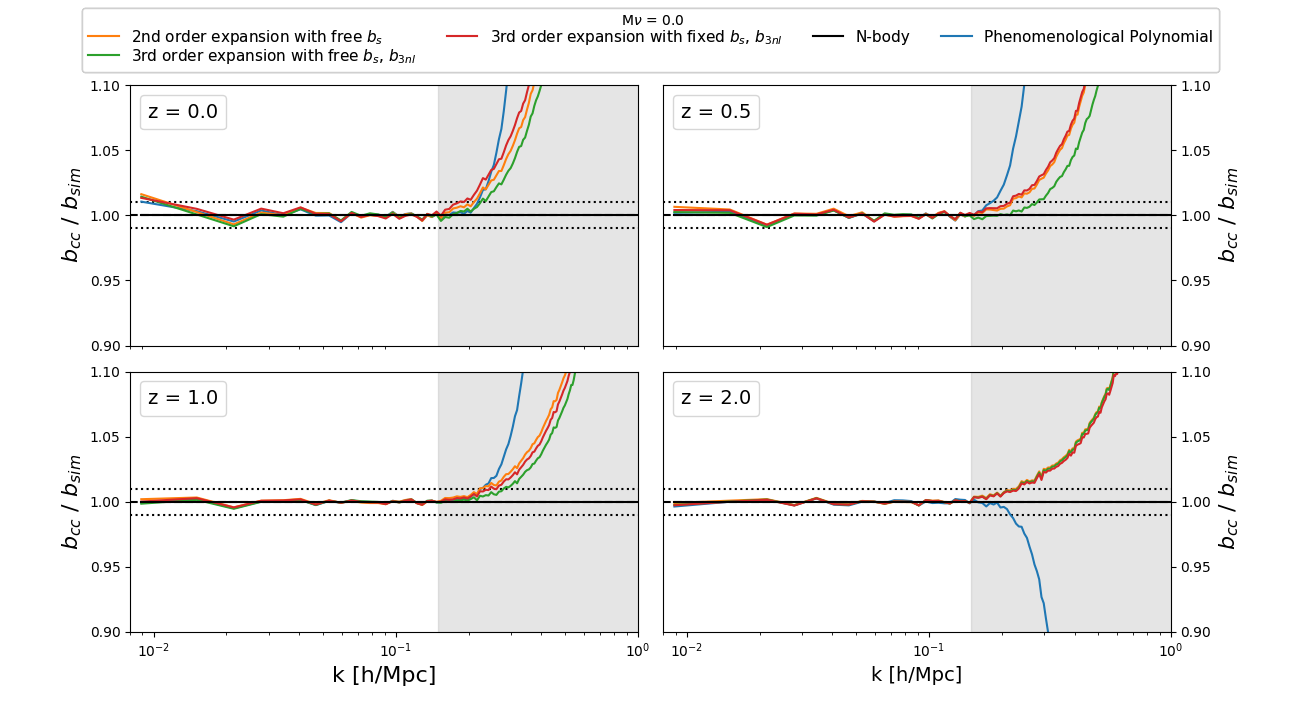}
\caption{The top panels show the halo bias from the simulations with massless neutrinos for different mass ranges (colored lines) at different redshifts (different panels). We fit these results with the  perturbation theory-base model for halo bias  (see Eq. \ref{bpt1}) up to $k_{\rm max}=0.15$ h/Mpc. We show the best-fits for the models with second-order bias (dashed), third order bias with $b_{s2}$ and $b_{3nl}$ as free parameters (dotted) and third order bias with $b_{s2}$ and $b_{3nl}$ fixed (dot-dashed). The black solid lines correspond to the polynomial model  fit  up to the same $k_{\rm max}$. The bottom four panels show the ratio between the best-fit models and the results of the simulations. For clarity, we only show the average ratio of the four mass ranges. The models based on perturbation theory work as well as the polynomial model Eq.~\ref{plodd} (within $1\%$ for the fitted k-range) but perform better on extrapolation beyond $k_{\rm max}$.}
\label{bcompare}
\end{center}
\end{figure*}
 
Eq. \ref{bpt1}, either at second or third order, represents thus our model for the halo power spectrum in configuration space. Note that thanks to  Eq.~\ref{eq:rescaling}, this perturbation theory-based model is only used for the massless neutrinos cases, which is where its validity and performance has been extensively tested. The halo bias $b_{\rm cc}$ is then obtained  from the ratio between $P_{\rm hh}(k)$ and $P_{\rm cc}(k)$, which we  fit  to  the N-body simulations with massless neutrinos for the different mass ranges and redshifts.  
We show the results in Fig.~\ref{bcompare} where we have set $k_{\rm max}=0.15$ h/Mpc  at all redshifts (case III). 

The different perturbation theory models reproduce, within $\simeq1\%$, the results of the simulations in all cases.  As predicted by Saito et al. \cite{39}, we also find that (although not easily apparent from the figures)  the model where $b_{s2}$ and $b_{3nl}$ are left as a free parameters performs slightly better than the model where they are fixed, in particular on large-scales. 

For comparison we also show the  polynomial bias model  fitted to  the same $k_{\rm max}$.  The models based on perturbation theory work as well as the polynomial model (within $1\%$ for the fitted k-range) but perform better on extrapolation beyond $k_{\rm max}$.
 
The best-fit values of the bias coefficients for the different perturbation theory models of this section, for all the mass bins  and redshift snapshots are reported in Appendix \ref{sec:tabelsbiascoeffs2}.

\subsection{Performance as a function of  $k_{\rm max}$ and discussion}
\label{max scale}

\begin{figure}
\makebox[\textwidth]{
\includegraphics[scale=0.5]{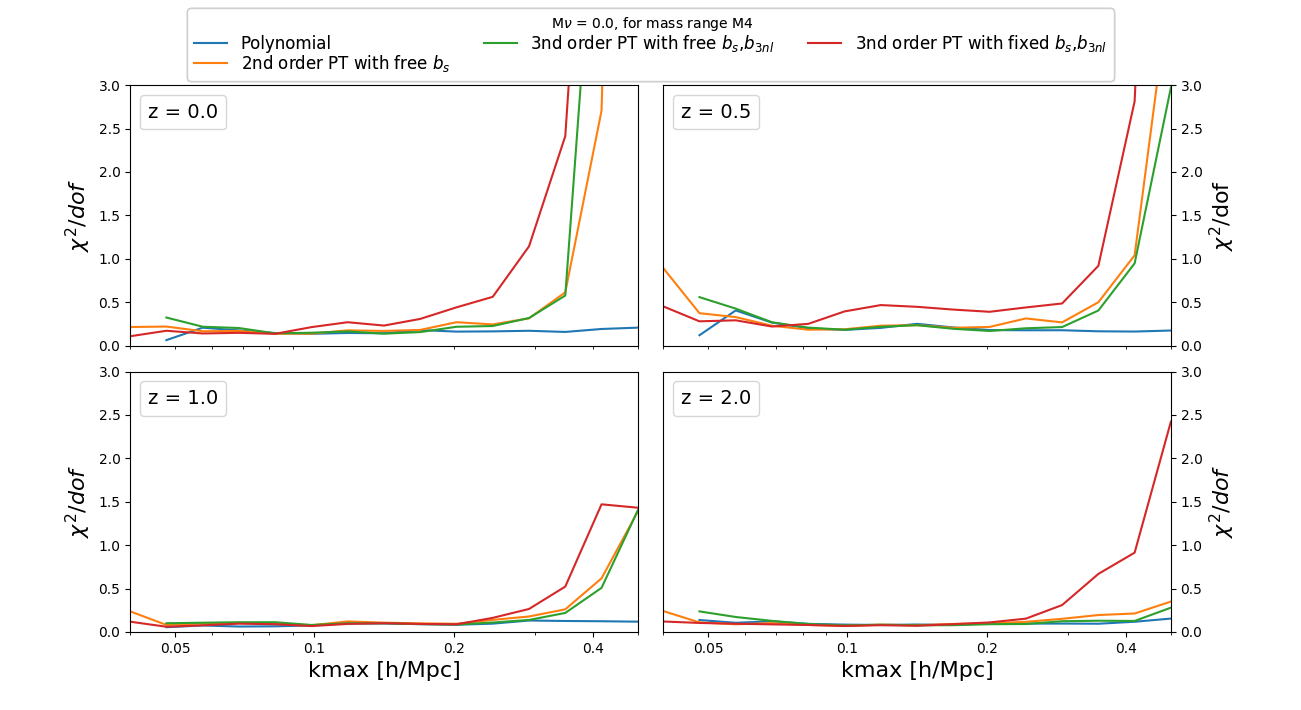}}
\caption{Reduced chi square, $\chi^2/{\rm dof}$, as a function of $k_{\rm max}$ for  the halo bias fit for  the mass range M4 (similar results hold for the other mass ranges) from the massless neutrino simulations:  polynomial  model (blue line) and the different perturbation theory models (orange, green and red; see legend) at different redshifts.  While the absolute $\chi^2$ amplitude is not meaningful, a sharp increase in $\chi^2$ with $k_{\rm max}$ denotes breakdown of the model.   In general, the perturbation theory model that performs better is the  third order bias with free $b_{s2}$ and $b_{3nl}$. 
}
\label{kmaxa}
\end{figure}

 Above we adopted $k_{\rm max} = 0.15$ h/Mpc when fitting the perturbation theory models to the results of the simulations,  finding excellent agreement. This is not surprising since these are mildly-non linear scales. Given the extra information present on smaller scales we explore performance of the model as a function of  $k_{\rm max}$.

  In Fig. \ref{kmaxa} we show the value of the reduced chi square, $\chi^2/{\rm dof}$, as a function of $k_{\rm max}$ at different redshifts for the massless neutrino case and for mass bin M4. Since in the fit we do not account for the correlations between $k$-bins, the absolute value of the $\chi^2$ is not meaningful, but relative values can be used to compare models.  As expected, perturbation theory works very well on large scales, but it fails on small scales: perturbation theory-based halo bias models breaks down at $k\sim\{0.15, 0.2,  0.25, 0.3\}$ h/Mpc at redshifts $z=\{0,0.5,1,2\}$. Of the perturbation theory based models, the one with more free parameters,  3rd order bias with free $b_{s2}$ and $b_{3nl}$, always performs better.
 For comparison we also show the performance of the  polynomial model. Very similar results hold for the other mass ranges.

\textbf{BE-HaPPy} implements both the polynomial and the perturbation theory models. The polynomial model is very accurate on small scales and very fast to evaluate, but its cosmology-dependent part is very approximate and the model itself is not physically well motivated. The perturbation theory models are on the other hand well motivated theoretically, correctly incorporate the dependence on cosmology but its range of validity is smaller than the polynomial model and is more computationally expensive to evaluate. Depending on the requirement of the problem, the user has the freedom to choose between the two approaches.

Inspection of the reported errors on the  best fit bias parameters in  Appendix \ref{sec:tabelsbiascoeffs1} and \ref{sec:tabelsbiascoeffs2}, indicates that the perturbative expansion coefficients are much better constrained than the polynomial fit  coefficients. Not surprisingly, the bias coefficient that more closely determines the large-scale bias is the best constrained parameter, with similar errors across the different models.

This is in large part because we report marginalised errors, and in the polynomial model the parameters  are much  more correlated than in the perturbation theory-based approach. The parameters of the perturbation theory-based approach are reasonably well constrained, even the third order bias. 
Our interpretation is that the parameters in the  perturbative   expansion are "physical" parameters and as such have well defined and roughly independent effects on the observables. While the coefficients in the polynomial expansion are effective parameters, which, taken  individually, do not correspond to a specific physical effect. As a result they are more correlated.   We thus conclude that the the perturbation theory approach represent a better "basis" to retrieve information on bias and cosmology.

We envision that these considerations  may  be useful even for application beyond  the scope of this paper.

\subsection{Massive neutrinos}
\label{large scale}

We now discuss  how to connect the real-space halo-bias  for the massless neutrino case to that in the massive neutrino case; in other words we estimate the performance of Eq.~\ref{eq:rescaling}.  In analyses where the overall bias amplitude is a nuisance parameter,  the correct calibration of $b^{\rm LS}_{\rm cc}$ becomes unimportant.

\begin{figure*}
\begin{center}
\includegraphics[scale=0.46]{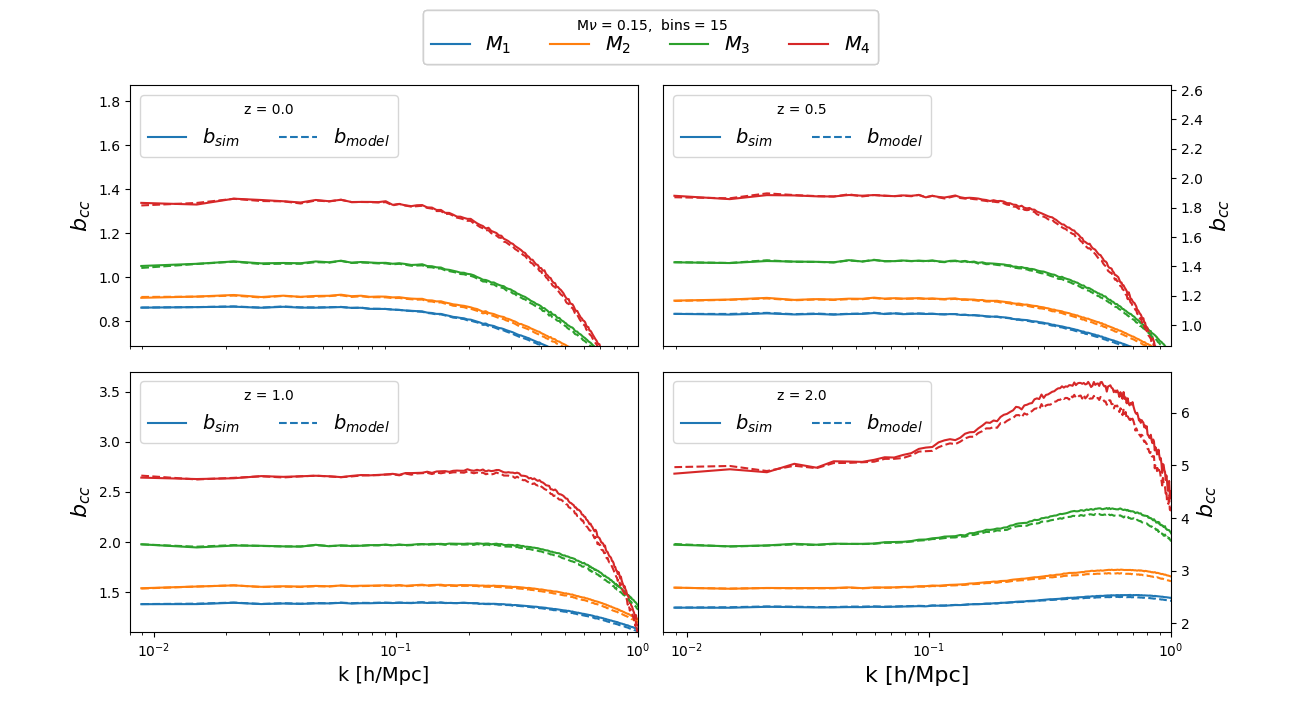}
\includegraphics[scale=0.46]{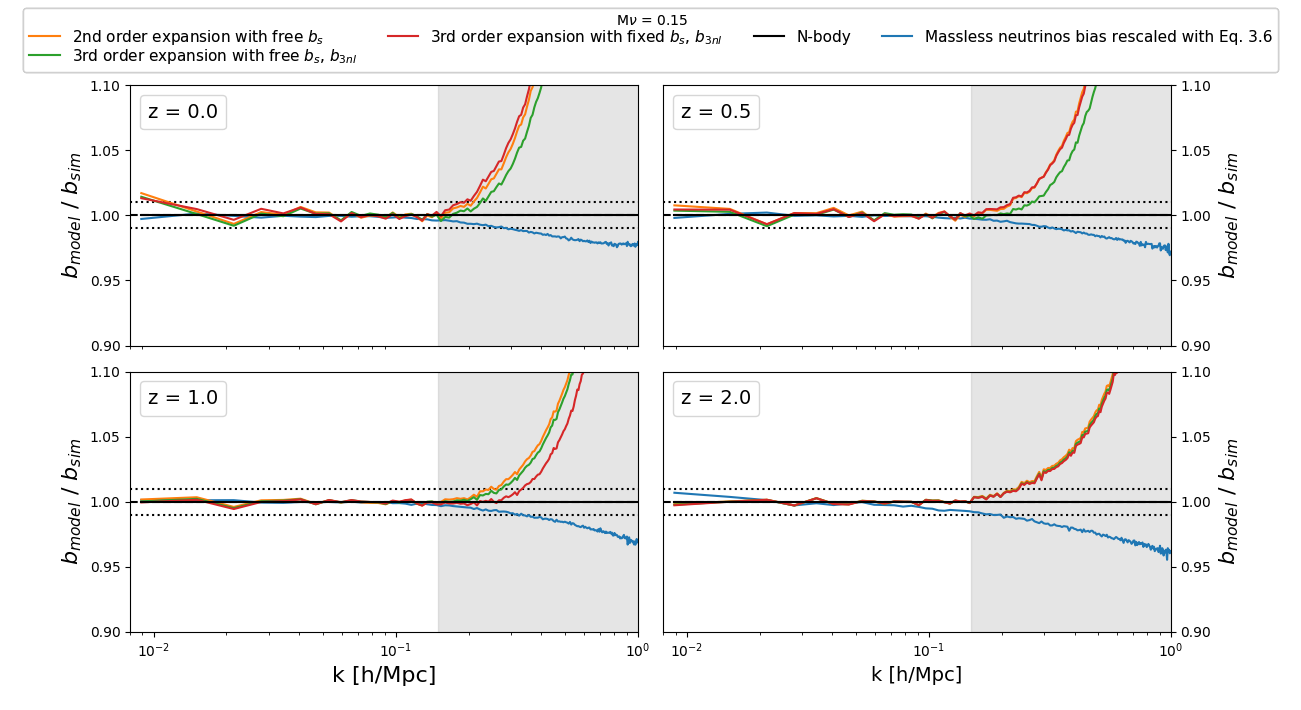}
\caption{The top panels show the halo bias of the massive neutrino model for different mass ranges at different redshifts. The solid lines represent the results of the N-body simulations, while the dashed lines correspond to our prediction through Eq. \ref{analytic_bias}. For clarity we do not show the scatter of the simulations results (they are very similar to those of the massless neutrino model). The bottom panels show the ratio between simulations outputs and the model fit. In all cases, we only fit up to $k_{\rm max}=0.15$ h/Mpc. The lines show the mean among the four different mass ranges. We find that our model to relate the bias of massive neutrino models to massless neutrinos models works very well down to the smallest scale we consider. Our perturbation theory model is also able to accurately describe the results of the simulations up to $k_{\rm max}$.}
\label{tinker3}
\end{center}
\end{figure*}

The approach of Eq.~\ref{eq:rescaling} requires  the value of the linear bias in the  massive neutrinos case. We will assume that analytical bias models, while not accurate enough to reproduce the linear bias from simulations at percent-level, can reproduce the {\em ratio} of the linear bias to the required accuracy:

\begin{equation}
\frac{b_{\rm cc}^{\rm LS}(M_{\nu} \neq 0)}{b_{\rm cc}^{\rm LS}(M_{\nu} =0)} = \frac{b_{\rm cc, model}^{\rm LS}, M_{\nu} \neq 0)}{b_{\rm cc, model}^{\rm LS}(M_{\nu} =0)}~,
\label{rule3}
\end{equation}
where $b_{\rm cc}$ refers to the value of the  simulations' bias (i.e., the  square root of the ratio   between $P_{\rm hh}$ and $P_{\rm cc}$)  while $b_{\rm cc, model}$ stands for the analytic value of the large-scale bias, which can be calculated as:
\begin{equation}
b_{\rm cc,model}^{\rm LS}(z, M_{\rm bin})= \frac{\int_{M_{\rm min}}^{M_{\rm max}}n(M,z)\:b(M,z)dM}{\int_{M_{\rm min}}^{M_{\rm max}}n(M,z)dM}~,
\label{bfid}
\end{equation}
where $n(M,z)$ and $b(M,z)$ are the analytic halo mass function and linear (scale-independent)  halo bias at redshift $z$ for haloes of mass $M$. The right-hand side of Eq. \ref{rule3} can then be computed numerically without running expensive simulations. In our calculations we have made use of the Crocce et al. halo mass function \citep{30} while we use the fitting formula of Tinker et al. \cite{29} to estimate the halo bias\footnote{The use of the Crocce mass function is motivated by the fact that   we find halos in our simulations with the friends-of-friends algorithm.  The Tinker mass  function is accurate for halos found via the  spherical overdensities method.  Note that we use the Tinker halo bias fitting formula, not the peak background-split halo bias derived from the Tinker mass function.  While in principle one could have derived the  peak background-split halo bias from the Crocce mass function, this has not been presented in the literature and it goes beyond the scope of this paper.}. We emphasize that in order to compute $b_{\rm cc, model}^{\rm LS}(M_{\nu})$ we have used the CDM+baryons power spectrum and not the total matter power spectrum.  We  show below that the above formula works very well. 

We can finally express the halo bias in models with massive neutrinos as a simple function of the halo bias in the model with massless neutrinos (see Eq.~\ref{eq:rescaling}):
\begin{equation}
b_{\rm cc} ( k,M_{\nu}) = b_{\rm cc} (k,M_\nu=0) \alpha_{\rm model}= b_{\rm cc} (k,M_\nu=0)\frac{ b_{\rm cc, model}^{\rm LS}(M_{\nu})}{ b_{\rm cc, model}^{\rm LS} (M_\nu=0)}~.
\label{analytic_bias}
\end{equation}

As discussed around  Eq.~\ref{eq:rescaling},  supported by Refs. \cite{1,2,4,5,6} here we assume that the definition of the bias with respect to cold dark matter + baryons removes all  scale dependence  due to neutrino mass on linear and mildly non-linear scales. As a consequence the only effect of massive neutrino is a change of overall  amplitude.  Values of the $\alpha$ coefficients as a function of mass bin and redshift are shown  in Appendix \ref{sec:alpha}.
In practice this means that all the bias coefficients ($b_1$, $b_2$, $b_{s2}$ and $b_{3nl}$) must  be rescaled by $\alpha_{\rm model}$ to achieve Eq.~ (\ref{analytic_bias})  for $b_{\rm cc}$. 
The above equation is expected to hold when the models with massive and massless neutrinos share the value of $\Omega_{\rm m}$, $\Omega_{\rm b}$, $h$, $n_s$ and $A_s$. In the top panels of Fig.~\ref{tinker3} we show with solid lines the halo bias of the massive neutrino model at different redshifts for different halo masses. The dashed lines in the top panels display our prediction using Eq.~\ref{analytic_bias}. As can be seen, the agreement is excellent in all cases; under $1\%$ for the scales of interest and below $5\%$ all the way to $k=1$ h/Mpc at $z=2$.

On small scales, for very massive haloes and at high-redshift some differences appear between the simulations and our rescaling procedure. This is  somewhat expected (see discussion after Eq.~\ref{eq:rescaling}) since massive haloes are highly biased (see appendix~\ref{eachmass} for a figure with  mass-bin dependence of the fit residuals.) Note that most deviations happen beyond the interesting k-range used for the fit, making this issue not too crucial.
However it is expected that  $\sigma_8$ will affect the bias coefficients. The massive and massless neutrinos simulations despite having the same $A_s$ have different $\sigma_8$. Interestingly,  Appendix \ref{sec:sigma8} shows that  large part of the effect is due to the different $\sigma_8$  between the massless and massive simulations, indicating that Eq.~\ref{eq:rescaling} holds when $\sigma_8$ is kept constant and not the primordial amplitude $A_s$. A detailed discussion  on this point has been presented in Refs \cite{s81,s82,s83}, therefore a more thorough discussion  goes beyond the scope of this paper.

\begin{figure}
\makebox[\textwidth]{
\includegraphics[scale=0.5]{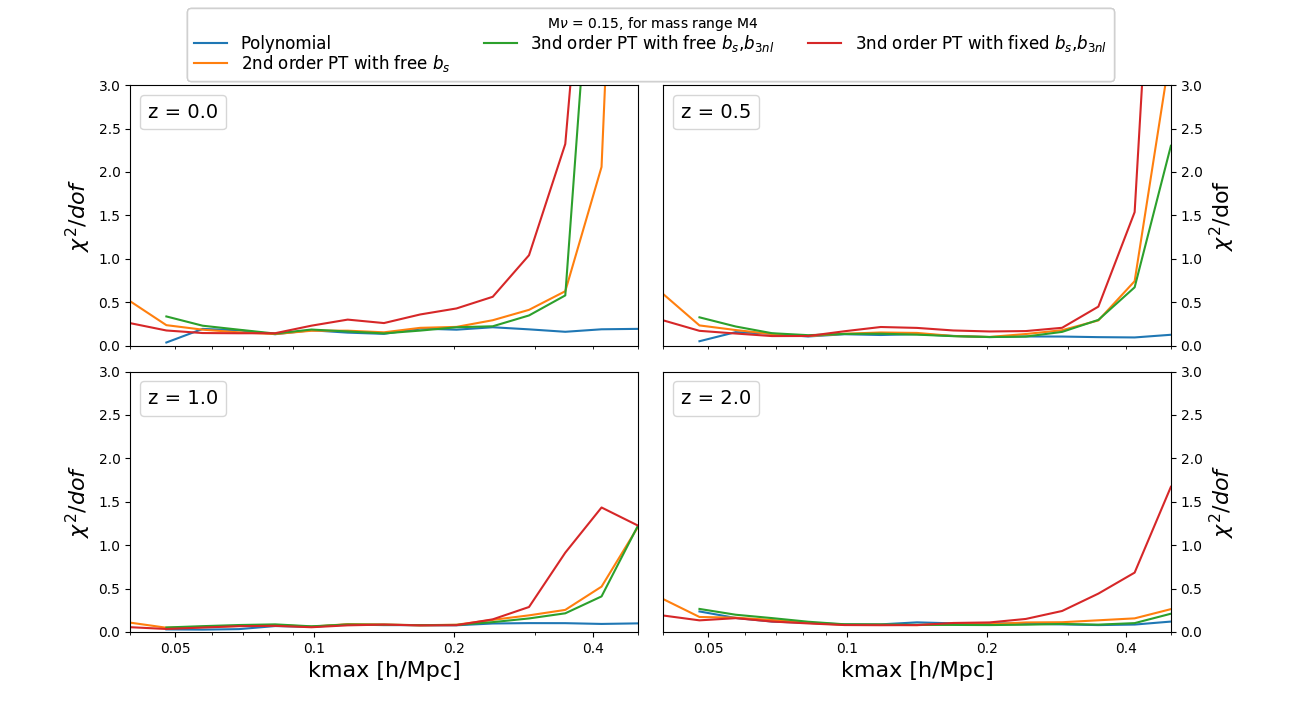}}
\caption{Reduced chi square, $\chi^2/{\rm dof}$, as a function of $k_{\rm max}$ for  the halo bias fit for  the mass range M4 (similar results hold for the other mass ranges) from the massive neutrino simulations:  polynomial  model (blue line) and the different perturbation theory models (orange, green and red; see legend) at different redshifts.  While the absolute $\chi^2$ normalisation is not meaningful, a sharp increase in $\chi^2$ with $k_{\rm max}$ denotes breakdown of the model.   In general, the perturbation theory model that performs better is the  third order bias with free $b_{s2}$ and $b_{3nl}$. 
}
\label{kmaxmassive}
\end{figure}

To highlight the accuracy of the fitting and rescaling procedure, we compare the massive neutrinos simulations' bias with our bias models in the bottom panels of Fig.~\ref{tinker3} (see caption for details).
 We find that these models are able to describe very accurately, $\simeq1\%$ level, the massive neutrinos simulations' outputs.  Similarly to what is shown in Fig.~\ref{kmaxa}, in Fig.~\ref{kmaxmassive} we show the value of the reduced chi square, $\chi^2$, as a function of $k_{\rm max}$ for the massive neutrinos case. The  sharp increase in $\chi^2$ with $k_{\rm max}$ denoting breakdown of the model happens at very similar scales as in Fig.~\ref{kmaxa} for the massless neutrinos case.
  
%%%%%%%%%%%%%%%%%%%%%%%%%%%%%%%%%%%%%%%%%%%%%%%%%
%%%%%%%%%%%%%%%%%%%%%%%%%%%%%%%%%%%%%%%%%%%%%%%%%
\section{Halo clustering in redshift-space}
\label{sec:RSE}

Peculiar velocities induce  clustering anisotropies along the line of sight called redshift-space distortions (RSD). RSD if accurately modelled,  can be used to to retrieve cosmological information, such as the growth rate of matter perturbations. Here we build on \citep{41,49} to model the effects of massive neutrinos on halo bias in redshift-space.

Below we present  the four different models we consider to describe redshift-space distortions before quantifying the accuracy of our models against the  simulations' results.

\subsection{RSD model I: Linear theory, Kaiser}
\label{sec:kaiser}
Villaescusa-Navarro et al. 2018 \cite{41} showed that, at linear level --Kaiser RSD~\cite{43}--, the (shot-noise subtracted) halo power spectrum in redshift-space (indicated by the $s$ superscript) in models with massive and massless neutrinos is given by
\begin{equation}
P^s_{\rm hh}(k,\mu) = (b_{\rm cc} + f_{\rm cc}\: \mu^2)^2  P_{\rm cc}(k) ~,
\label{kaiser}
\end{equation}
where $P_{\rm cc}(k)$ is the CDM+baryons power spectrum,  $\mu$ the  cosine of the angle with respect to the line of sight and $f_{\rm cc}$ is the linear CDM+baryons growth rate ($d\ln \sqrt{P_{\rm cc}(k,a)}/d\ln a$, with $a$ the scale factor) of the CDM+baryons component.  While in the original Kaiser \cite{43} formulation the configuration space power spectrum should be the linear one, in what follows we will use  a non-linear $P_{\rm cc}(k)$ in Eq.~\ref{kaiser} as well as the full scale-dependent $b_{\rm cc}(k)$. In the figures below we will use  $P_{\delta\delta}$ computed with FAST-PT (see Appendix \ref{sec:redshift-space}) from a CDM + baryons linear power spectrum for $P_{\rm cc}(k)$ and  our model (see captions) for  $b_{\rm cc}(k)$.

\subsection{RSD model II: Scoccimarro}
\label{sec:sco}
Scoccimarro \cite{42} was among the firsts to propose a non-linear extension of the large-scale, linear Kaiser model for RSD: 
\begin{equation}
P^{s}(k,\mu) = P_{\delta\delta}(k) + 2f\mu^{2} P_{\delta \theta}(k) + f^{2}\mu^{4} P_{\theta \theta}(k) %\times {\rm exp}(-f^{2}k_{z}^{2}\sigma_{v}^{2})
\label{sco}
\end{equation}
where $P^s(k,\mu)$ is the matter power spectrum in redshift-space\footnote{In what follows in order not to carry too many subscripts when in redshift space and when not ambiguous we will drop the $mm$ subscript from the matter power spectrum symbol.},  $P_{\delta\delta}(k)$, $P_{\delta \theta}(k)$ and $P_{\theta \theta}(k)$ are the density, density-velocity and velocity power spectrum, respectively.  

For our purposes,  in models with massive neutrinos, the density and velocities  and therefore the power spectra are the ones of the CDM+baryons field, not of the total matter field.  $P_{\delta \theta}$ and $P_{\theta \theta}$ have the same shape as $P_{\delta\delta}$, and thus can be computed with FAST-PT in the same fashion. As an example we take Eq. 64 of \cite{42}:
\begin{equation}
P_{\theta\theta}(k) = \underbrace{P(k)}_{\text{linear part}} + \underbrace{2\int [G_{2}(p,q)]^{2}P(p)P(q)d^{3}q}_{\text{$P_{22}$(k)}} + \underbrace{6P(k)\int G_{3}(k,q)P(q)d^{3}q}_{\text{$P_{33}$(k)}}
\end{equation}
where $G_{2}(p,q)$ and $G_{3}(k,q)$ are perturbation theory kernels. The $P_{22}$ convolution integrals are computed using spherical harmonics after the kernel is expanded in Legendre polynomials (cf. section 2.2 of \cite{8}). The $P_{13}$ integrals are more difficult because the wavenumber structure is different and the kernels are more complicated. It also requires regularization to correct for IR divergence (see Sec.~2.3 and 2.4 of \cite{8}). As a cross check of our implementation, we compare the results of Scoccimarro et al. \cite{42} with our  calculations, obtained using similar cosmological parameters to theirs, finding a good agreement (see Fig. \ref{scocci} in Appendix \ref{sec:redshift-space}). 
To generalise Eq.~\ref{sco} to haloes we use
\begin{equation}
P^{s}_{\rm hh}(k,\mu) = b_{\rm cc}^2(k)P_{\delta\delta}(k) + 2b_{\rm cc}(k)f\mu^{2} P_{\delta \theta}(k) + f^{2}\mu^{4} P_{\theta \theta}(k)  \,. %\times {\rm exp}(-f^{2}k_{z}^{2}\sigma_{v}^{2})
\label{scohalos}
\end{equation}

\subsection{RSD model III: TNS}
\label{tns}
The above model is the basis for one of the most popular models of redshift-space distortions: the Taruya, Nishimichi and Saito (TNS) model \cite{44} where several coefficients were added to the Scoccimarro model to account for the mode coupling between the density and velocity fields. In summary TNS adds two ``coefficients'', $A$ and $B$,  that depend on $k$, $\mu$ and $f$ to Eq.~\ref{sco}.

FAST-PT incorporates routines to compute these coefficients for the matter power spectrum. If we apply a linear bias to the matter fluctuation $ \delta_{g} \longrightarrow b_1\delta(x)$, it is easy to show that  expressions for the A and B coefficients for the halo power spectrum become\footnote{A and B are in fact proportional to $b^2$, the other powers of $b$ come from the $k\mu f$ factor in the integrals of $A(k,\mu,f)$ and $B(k,\mu,f)$.}:
\begin{equation}
A(k,\mu,f) \Rightarrow b_1^3A(k,\mu,\beta)
\label{A}
\end{equation}

\begin{equation}
B(k,\mu,f) \Rightarrow b_1^4B(k,\mu,\beta)
\label{B}
\end{equation}

 where $b_{1}$ is the linear bias, $\beta = f/b_{1}$.
While a linear bias  approximation is not sufficient for this model, as indicated in \cite{44, 37, 39} the bias  coefficient in front of the $A$ and $B$ functions, which are higher-order corrections, is the linear one i.e., $b_1$.

\subsection{RSD model IV: eTNS} 
\label{sec:etns}
To go beyond linear bias, we consider the so-called eTNS bias model \cite{37,39}. 
\begin{equation}
\begin{split}
P_{\rm hh}^s(k, \mu) &= P_{\rm hh}(k) + 2f\mu^{2} P_{h,\delta \theta}(k) + f^{2}\mu^{4} P_{\theta \theta}(k) + b_{1}^{3}A(k,\mu,\beta) 
+ b_{1}^{4}B(k,\mu,\beta) \\
%&\times {\rm exp}(-f^{2}k_{z}^{2}\sigma_{v}^{2}\mu^{2})~,
\end{split}
\label{etns}
\end{equation}
where $b_{1}$ is the linear bias, $\beta = f/b_{1}$ and it is assumed that there is no velocity bias. $P_{\rm hh}(k)$ is given by Eq. \ref{bpt1} and the expression of $P_{\rm h,\delta\theta}(k)$ is given by \cite{37}\\
\begin{equation}
P_{\rm h,\delta\theta}(k) = b_{1}P_{\delta\theta}(k) + b_{2}P_{b2,\delta}(k)
+ b_{s2}P_{bs2,\delta}(k) + b_{3nl}\:\sigma_{3}^{2}\:P_{\rm}^{\rm lin}(k) \,.
\label{pdt}
\end{equation}
We limit ourselves to the linear bias term when computing the $A$ and $B$ correction terms in Eq. \ref{etns}. Of course, for models with massive neutrinos, the above quantities need to be computed by using the CDM+baryons power spectrum, not the total matter power spectrum.

\subsection{Fingers of God}
\label{sec:fog}
The motions of particles/galaxies inside haloes induce a characteristic feature in redshift-space: the so-called Fingers-of-God (FoG). When modeling redshift-space distortions, it is important to account for this effect, as it dominates the amplitude and shape of the power spectrum on small scales but can also propagate to large scales. 

Here we  characterize the FoG as: 

\begin{equation}
F(k,\mu) = \exp\left[-k^2 f(z) \sigma_v(z)^2\mu^2\right]
\label{eq:fog}
\end{equation}
where $\sigma_v(z)=D(z)\sigma_0$, $D(z)$ is the linear growth rate of perturbations normalised to unite at $z=0$  and $\sigma_0$ is a free parameter representing the effective velocity dispersion of particles/galaxies inside halos. This approach  goes under the "streaming" models category i.e., the FoG term is treated independently of the linear and mildly non-linear effects. 
The effect of FoG on the clustering of haloes should be 
small if not negligible, but \textbf{BE-HaPPy}  allows the user to optionally include it\footnote{ A small but possibly non-negligible value for $\sigma_0$ for halos has been found in the literature before \cite{67}. It can be argued that this extra degree of freedom absorbs high $k$ residuals  in the fit arising from limitations of the modelling.}.

\begin{figure}
\begin{center}
\includegraphics[scale=0.485]{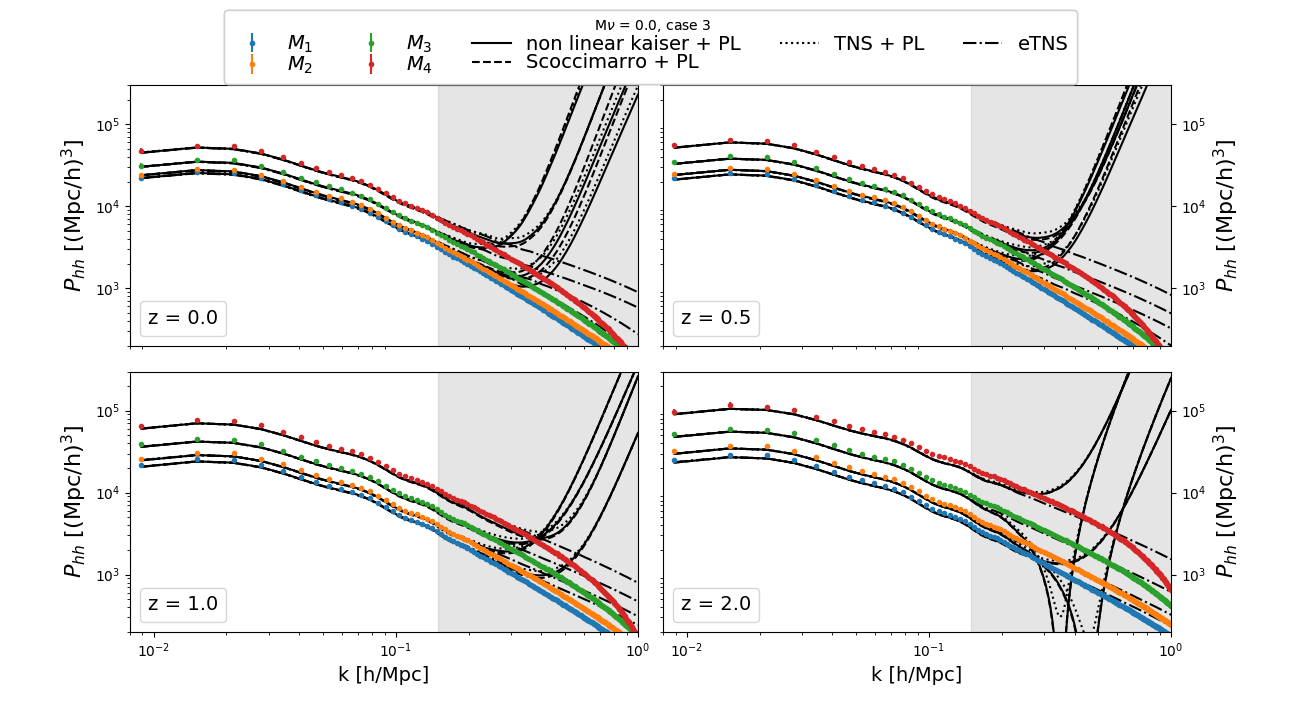}
\includegraphics[scale=0.485]{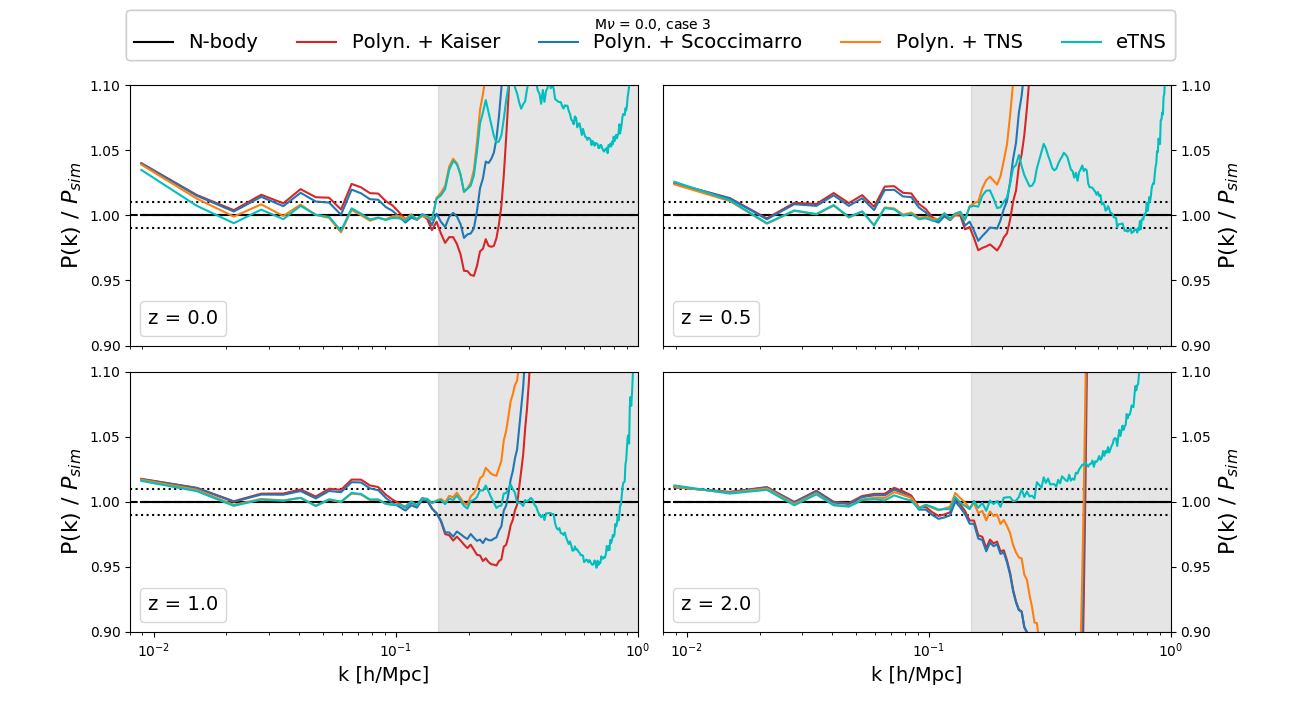}
\caption{The top panels show the  massless neutrinos  N-body simulations'  redshift-space power spectrum monopole  for different mass ranges at different redshifts. We then fit these results using our four different theoretical models: non-linear Kaiser \ref{kaiser} with polynomial bias (solid black), Scoccimarro \ref{sco} with polynomial bias (dashed black), TNS \ref{tns} with polynomial bias (dotted black) and eTNS \ref{etns} (dot-dashed black). In all cases we set $k_{\rm max}=0.15$ h/Mpc for the fit (case III). The bottom panels display the ratio between the fits and  the simulations outputs. For clarity, we show the average results of  the four different mass ranges. The  models reproduce accurately the results of the simulations, with the eTNS performing better in all cases.}
\label{red_massless1}
\end{center}
\end{figure}

\subsection{Comparison to massless neutrinos simulations}
\label{sec:calib}
From the massless neutrino simulations, we  computed the monopole of the  halo redshift-space power spectrum  for different mass ranges at different redshifts. To improve the statistics, we have  taken the average of  RSD along the three cartesian axes.  The simulation outputs are used to study the accuracy of our theoretical models: the Kaiser (Sec.~\ref{kaiser}, with non-linear rather than linear power spectrum-- hereafter non-linear Kaiser), Scoccimarro (Sec.~\ref{sco}), TNS (Sec.~\ref{tns}) and eTNS (Sec.~\ref{etns}).

\begin{figure}
\makebox[\textwidth]{
\includegraphics[scale=0.5]{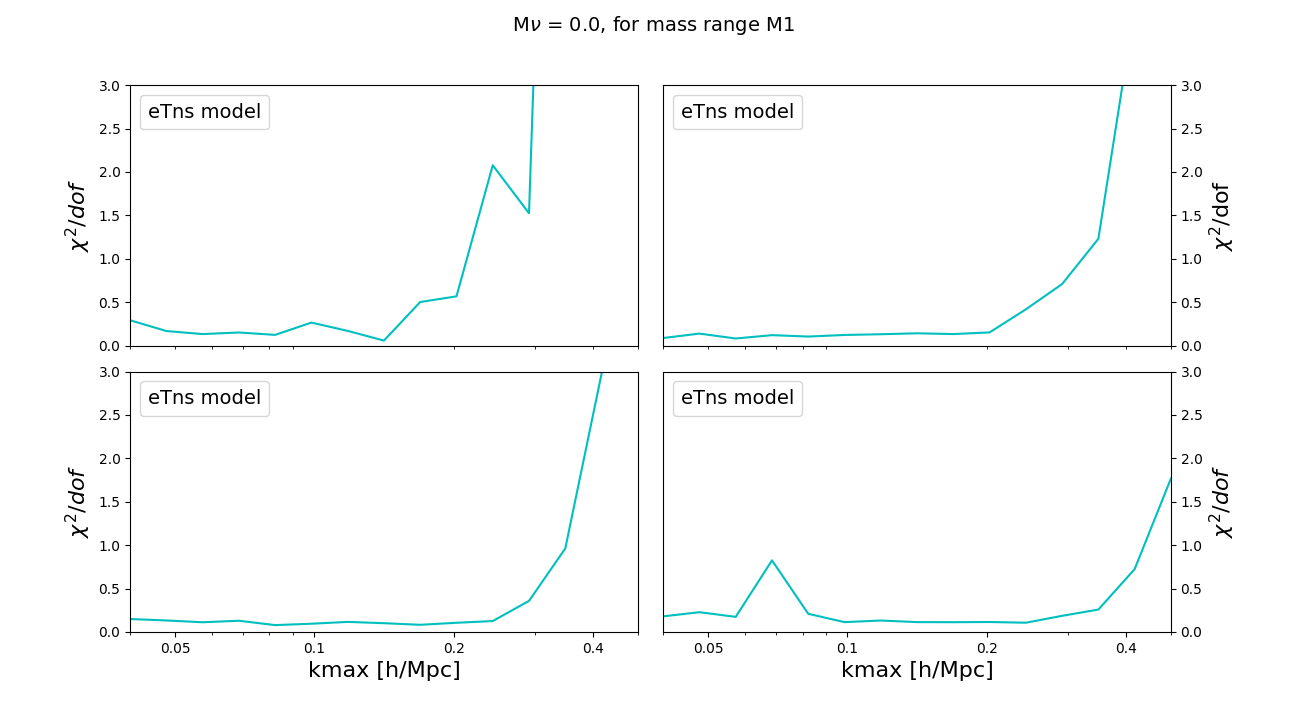}}
\caption{Reduced chi square, $\chi^2/{\rm dof}$, as a function of $k_{\rm max}$ for  the halo power spectrum fit for  the mass range M1 (similar results hold for the other mass ranges) from the massless neutrino simulations in redshift-space:  eTNS model at different redshifts.  While the absolute $\chi^2$ normalisation is not meaningful, a sharp increase in $\chi^2$ with $k_{\rm max}$ denotes breakdown of the model.  
 }
\label{kmaxb}
\end{figure}

For each of the above  models, we fit the massless neutrinos simulations'  redshift-space power spectrum monopole   at different redshifts and for different mass ranges and $k_{\rm max}=0.15$ h/Mpc. In this case we do not fit for the value of the bias parameters, but use the values we obtained from  the real space fit. Thus, the only free parameter is $\sigma_0$.  We show the results in Fig. \ref{red_massless1}.  The best fit $\sigma_0$ is in the range $6-8$ Mpc/h which is consistent with the findings of e.g.,\cite{67}.

All models are able to describe the clustering of haloes in redshift-space \LV{as observed in the HADES simulations}  up to $k_{\rm max}$, with percent accuracy. As expected, the model that performs better is eTNS, with sub-percent accuracy on  a wide  k-range.
The deviations of the model at the largest scales (top left panel of the bottom half of Fig.~\ref{red_massless1}) are probably due to sample variance.  In fact, to reproduce more closely a realistic analysis, we have used the  FAST-PT non-linear $P_{\rm cc}(k)$, instead of the simulations outputs. We have checked that using the simulations outputs  for $P_{\rm cc}(k)$ instead, the deviation  effectively disappears.

To qualitatively assess the  (small scales) breakdown of the modelling in redshift-space, in Fig.~\ref{kmaxb} we show the  reduced chi square, $\chi^2/{\rm dof}$, as a function of $k_{\rm max}$ for the eTNS model. The behaviour (the $\chi^2$ dependence on the the redshift and different bias models but weak dependence on the mass bins is expected from perturbation theory.).

\subsection{Comparison to Massive neutrinos simulations}
\label{sec:validation}

We finally quantify the performance of our  approach for the redshift-space massive neutrinos case.
The massless to massive neutrinos models re-scaling  Eqs.~\ref{eq:rescaling} and \ref{rule3} for the  Kaiser Eq.~(\ref{kaiser}) and Scoccimaro  Eq.~(\ref{sco}) models is straightforward, since the full expression of the bias $b_{\rm cc} (k)$ appear explicitly in the equations.  
For the other models,  all the perturbation theory bias coefficients $b_1, b_2, b_{s2}$ etc. (calibrated in real space and for massless neutrinos) must be rescaled according to Eq.~\ref{analytic_bias} and  related discussion. 

\begin{figure}
\begin{center}
\includegraphics[scale=0.49]{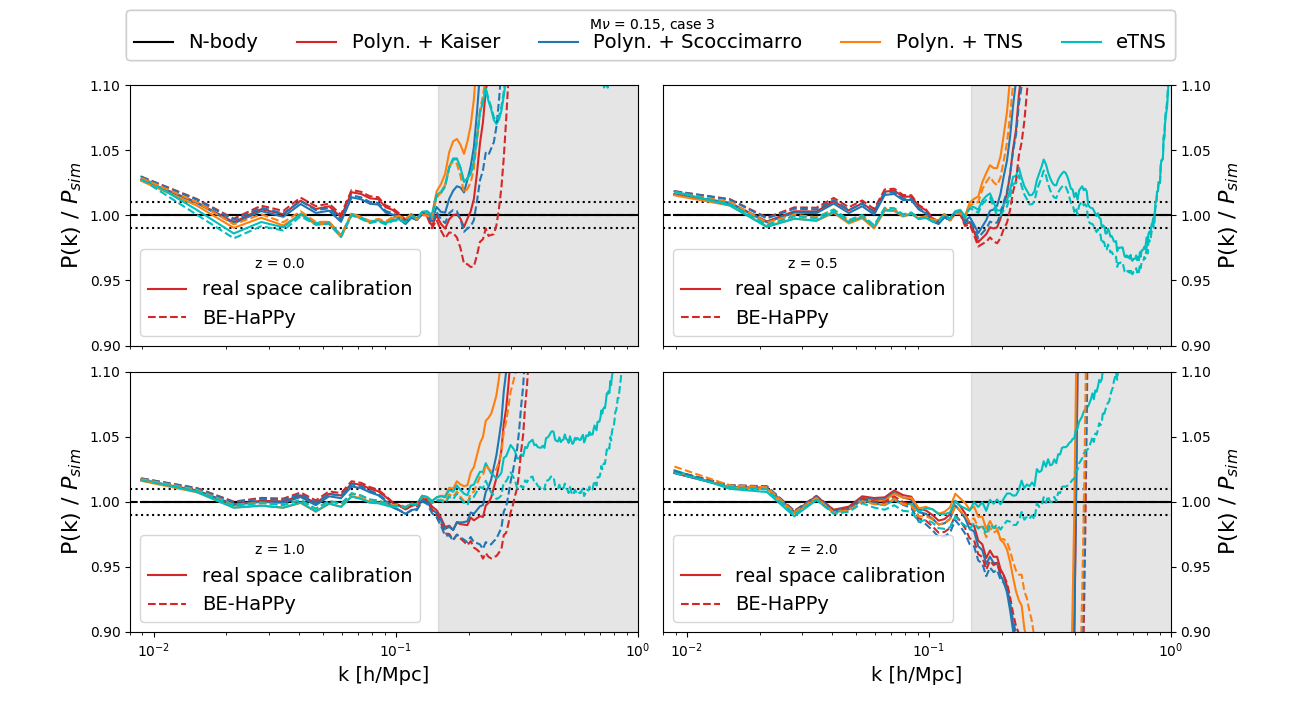}
\caption{Residuals between $b_{\rm cc}(k)$ in redshift-space from  \textbf{BE-HaPPy} and  $b_{\rm cc}(k)$ calibration done  for  (configuration space) massive neutrinos simulations. See Tab.~\ref{analytic} for details. The $> 1\%$ deviations at high $k$  and high $z$  arise from  the $\sigma_8$ mis match, see Appendix \ref{sec:sigma8}.}
\label{red_massive1}
\end{center}
\end{figure}

We use the halo bias $b_{\rm cc}$ model (or alternatively the $\alpha_{\rm model}$ for the  bias parameters in TNS and eTNS) calibrated in real space for massless neutrinos, leave $\sigma_0$ as a  free  parameter (in the spirit that in any analysis it will be a nuisance parameter to be   marginalised over)  re-scale the bias coefficients in the presence of massive neutrinos (eq.~\ref{analytic_bias}) and apply the redshift-space mapping of Sec.~\ref{sec:kaiser}, \ref{sec:sco}, \ref{tns}, \ref{sec:etns} with the FoG modelling of Sec.~\ref{sec:fog}.   To quantify the performance of this approach (calibration on massless neutrino simulations and rescaling) we compare this (benchmark) to a fit  to $b_{\rm cc}$ (or the perturbation theory  bias parameters) done directly on the massive neutrino simulations outputs (in real space).  
A summary  is reported in Tab.~\ref{analytic}. The comparison is shown in Fig.~\ref{red_massive1}.

%This is then benchmarked against a calibration done directly  for the massive neutrinos simulations and the comparison is shown in Fig.~\ref{red_massive1}.

\begin{table}[h!]
\centering
\resizebox{\textwidth}{!}{%
\begin{tabular}{|c|c|c|c|}
\hline
 & fit  of  $M_{\nu}\ne 0$ sims&  \textbf{BE-HaPPy} \\ \hline
bias coefficients & fitted on $b_{\rm cc} (k, M_\nu = 0.15)$  & fitted on $b_{\rm cc} (k, M_\nu = 0.0)$ \\ \hline
\begin{tabular}[c]{@{}c@{}}linear bias $b_1$ input for \\ A \ref{A} and B \ref{B} coefficients\end{tabular} & fitted on $b_{\rm cc} (k, M_\nu = 0.15)$ &  fitted on $b_{\rm cc} (k, M_\nu = 0.0)$ \\ \hline
velocity dispersion & free parameter  & free parameter \\ \hline
rescaling & no &  yes \\ \hline
\end{tabular}%
}
\caption{Benchmark  of Be-HaPPY performance  in Fig.~\ref{red_massive1}. BE-HaPPy  (third column) uses  only massless neutrino simulations to calibrate the fit and obtains  $b_{\rm cc}$  via rescaling. The performance of this is quantified by comparing it to a $b_{\rm cc}$  fit  done  on massive neutrino simulations (second column).}
\label{analytic}
\end{table}
 
The performance is qualitatively similar to that of the massless neutrinos case except for the highest redshift panel. The $> 1\%$ mis match at high $k$  and high $z$  arises from  the $\sigma_8$ mis match (See  discussion below Eq.~\ref{analytic_bias} and  Appendix \ref{sec:sigma8}).  

%%%%%%%%%%%%%%%%%%%%%%%%%%%%%%%%%%%%%%%%%%%%%%%%%
%%%%%%%%%%%%%%%%%%%%%%%%%%%%%%%%%%%%%%%%%%%%%%%%%
%%%%%%%%%%%%%%%%%%%%%% BE-HaPPy %%%%%%%%%%%%%%%%%%%%%
\section{BE-HaPPy}
\label{sec:BEH}
\textbf{BE-HaPPY}, that stands for Bias Emulator for Halo Power spectrum Python is a plug-in designed to be implemented in MCMC softwares\footnote{Our current implement supports only MontePython \cite{48}}. The primary goal of \textbf{BE-HaPPY} is to accurately predict the halo power spectrum in real- and redshift-space in a very computationally efficient manner. Explanations on the installation of the code, its usage and its various components are available on the author github account\footnote{\url{https://github.com/Valcin/BE_HaPPy}.The code in emulator  mode and calibration mode  will be made available on the same repository.  In this modality the code relies on our extension of the FAST-PT software, so any  public release must be coordinated across  different collaborations.}. 

\textbf{BE-HaPPY} as a plug-in for MontePython \cite{48} can be ran in two modes: \textbf{calibration mode} and \textbf{emulator mode}.

In the \textbf{calibration mode} the code  goes through all the calibration steps described in this paper. It provides our modified FAST-PT  and the calibration procedure. The user must supply  the necessary simulations outputs. In this way cosmologies different from the fiducial one used here (and different mass ranges,  redshifts snapshots etc.)  can be explored. While we expect that for cosmologies consistent with current data, and for the expected precision of  forthcoming surveys the provided calibration is good enough , it may be of interest to explore detailed dependence of calibration for other cosmological models to adjust to the required precision of next generation surveys.

The  good performance of the rescaling Eq.~\ref{analytic_bias} implies that only calibration on massless neutrinos simulations is really necessary, provided that   the corresponding massive neutrinos case of interest has the same values for the  other cosmological parameters and  in particular the same value for the  $\sigma_8$ parameter. Of course,  if the user envisions marginalising over the overall bias amplitude, calibration is also only necessary on massless neutrinos simulations. However, as long as the other cosmological parameters do not vary much, we expect our modelling to still perform well.   

In the \textbf{emulator mode},  \textbf{BE-HaPPY} uses the templates, bias coefficients and RSD modelling  calibrated for our fiducial cosmology  (or user supplied as a result of  a previous \textbf{calibration mode}  run) to provide an emulator for  the halo bias. This is then  used in the standard MCMC run. The implicit approximation done here is that in the  MCMC exploration of parameter space, the cosmology  does not  deviates too drastically from the fiducial one as to invalidate the calibration. This is more of a concern for the polynomial bias modelling than for the perturbation theory-based one. Note that marginalisation over bias parameters (with user-supplied priors)  is also an option of the code, thus making this mode (emulator+ bias parameters marginalisation) very robust to the choice of cosmology. Although beyond the scope of this paper, one could envision  sampling the  (cosmological) parameter space also  for other parameters than neutrino mass and use  techniques such as Gaussian processes to extend our modelling beyond the fiducial cosmology adopted here. \textbf{BE-HaPPY} would still provide the workhorse for such an effort. It could provide calibration in several  regions in parameter space around  specific sets of cosmological parameters. The the  Gaussian processes approach (or similar)  would smoothly  interpolate across these regions.
  
Below we summarise the features of  \textbf{BE-HaPPY}, more info can be found in the  code repository.

\begin{enumerate}
\item Four cumulative mass bins are available (see Table \ref{mr}).
\item Results for four different $k_{\rm max}$ values; cases I, II and III (see Table \ref{mr2}).
\item Outputs between $z=0$ and $z=2$; calibrations at redshifts 0, 0.5, 1 and 2 and interpolations in between.
\item Three models of bias are available: 1) linear, 2) polynomial (Sec.~\ref{plodd}), and 3) perturbation theory expansion up to third order (Sec.~\ref{bpt1}). 
\item Four RSD models are available: (non-linear) Kaiser (Sec.~\ref{sec:kaiser}), Scoccimarro (Sec.~\ref{sec:sco}), TNS (Sec.~\ref{tns}) and eTNS (Sec.~\ref{sec:etns}).
\item The user has the option to include the Fingers of God term (Sec.~\ref{eq:fog}) with $\sigma_{0}$ as a free parameter.
\item Text files of bias coefficients and PT terms. User has the option to  substitute these with those for  a different model/calibration.
\end{enumerate}

\textbf{BE-HaPPy} is designed to work with both models with massless and massive neutrinos. Importantly, the output for the massive neutrinos models is obtained through our proposed rescaling of the bias for massless neutrinos. This reduces the dimensionality of the parameter space, enabling a faster calculation. \textbf{BE-HaPPy} allows the user to output halo bias results with respect to CDM+baryons or total matter (Eq.~\ref{bmm2}).

We calibrated the emulator so \textbf{BE-HaPPy} achieves  percent or sub-percent precision on the scales of interest (see Figs. \ref{bcompare}, \ref{tinker3} (bottom panel), \ref{red_massless1} and \ref{red_massive1} (bottom panel)). This is the precision level achieved in fitting the relevant quantities from the HADES simulations. This is not necessarily the accuracy level achieved in fitting the relevant quantities in the real Universe.
Moreover  this calibration may not work as well for cosmologies that differ significantly from our fiducial one  and used in the simulations. 

We designed the code to be as modular  as possible,  providing text files for  the required quantities (bias and perturbation theory coefficients).
While we use FAST-PT to compute the non linear density spectrum $P_{\rm cc}(k)$  this can be substituted by  another cosmic emulator (e.g.,\cite{76}) or Halofit \cite{77}. It is also possible to  use softwares like \textbf{RelicFast} to include large-scales  linear effects not included here. \textbf{BE-HaPPy} may also be used with a different cosmological model as a test.  
To keep track of the impact of any  deviations from our settings,  we added an "error" feature in the code where the user can access the relative error (value and percentage) at each $k$ of the selected arrays between the power spectra computed by \textbf{BE-HaPPy} and those  obtained from the original suite of N-body simulations we used for the calibration. This feature is only available for the cosmology and neutrino masses ($M_\nu=0$ and $M_\nu=0.15$ eV) models of the simulations considered here.

%%%%%%%%%%%%%%%%%%%%%%%%%%%%%%%%%%%%%%%%%%%%%%%%%
%%%%%%%%%%%%%%%%%%%%%%%%%%%%%%%%%%%%%%%%%%%%%%%%%
%%%%%%%%%%%%%%%%%% CONCLUSIONS %%%%%%%%%%%%%%%%%%%%%%
\section{Conclusions}
\label{sec:conclusion}

We have presented  fast and accurate modelling of the halo bias in Fourier space which includes the effect of massive neutrinos and applies to both real and redshift-space. The modelling has been calibrated on  a suite of state-of-the-art N-body simulations (the HADES simulations). 

Our approach  relies on the fact that, unlike that defined with respect to the total matter, the halo bias with respect to the CDM+baryons, $b_{\rm cc}$, does not  show extra scale dependence induced by --and dependent on--   neutrino masses. Hence we have provided a detailed calibration  and analytic expression of $b_{\rm cc}(k)$ which holds into the mildly non-linear and even non-linear regime. We have used two approaches: one phenomenological, where the halo bias takes   a polynomial  form in $k$, and a perturbation-theory based.

The $b_{\rm cc}(k)$ model so calibrated on massless neutrinos simulations can then be converted  to that for massive neutrinos models by a simple (analytic) amplitude rescaling.  While we have carefully  quantified  how this rescaling works, and tested its performance  with massive neutrinos N-body simulations, it is important to keep in mind that in most cosmological analyses the (scale-independent) bias amplitude is treated as a nuisance parameter and marginalised over. The scale dependence of the halo bias however is important and must be accurately modelled as it has been shown that if neglected can induce statistically significant systematic shifts in the recovered cosmological parameters from forthcoming surveys.

The polynomial bias model reaches percent to sub-percent accuracy into the non-linear regime, the perturbation theory based model achieves the same  accuracy only in the mildly non-linear regime.  The modelling of redshift-space distortions, being also  perturbation theory-based, reaches  percent to sub-percent accuracy in the mildly non-linear regime. This is the accuracy level  at which the relevant quantities of the  input simulations are being recovered by \textbf{BE-HaPPy}. This reported accuracy level does not take into account that  the input simulations may not be a sub-percent description of the Universe. For example only specific halo mass bins were considered, the mass resolution of the simulations is set  as well as the fiducial cosmology. With the advent of more accurate  simulations  \textbf{BE-HaPPy} should be re-calibrated.

Observable tracers  such galaxies are are likely to  reside in dark matter halos, so while the  model we provide here  for the halo bias might not be sufficient to interpret future galaxy surveys, it is a necessary preliminary ingredient.
 
We provide a fast emulator for the halo bias (\textbf{BE-HaPPy}).   \textbf{BE-HaPPy} returns the  halo bias as function of  scale, redshift and  halo mass, in real or redshift-space for both massless and massive neutrino cosmologies, as well as the perturbation theory-based non-linear redshift-space halo power spectrum.  The user can select which modelling to use, the scales of interest and other option about e.g., redshift-space distortions implementation.  \textbf{BE-HaPPy} is fast enough to be included in standard Markov chain Monte Carlo runs   at only small additional computational cost. Since we have calibrated \textbf{BE-HaPPy} on a concordance $\Lambda$CDM set of cosmological parameters, the polynomial bias model might be less robust to change of cosmology than the perturbation theory approach. For cosmological models significantly different from the concordance $\Lambda$CDM we recommend the users to  check the \textbf{BE-HaPPy} performance and if needed to  re-calibrate it.

The next-generation large-scale  structure surveys will provide unprecedented wealth of information about the clustering properties of the Universe provided that the modelling  tools used reach the required accuracy. \textbf{BE-HaPPy} aspires to be one of them. It provides an easy solution to compute the halo power spectrum in massive and massless neutrinos cosmologies taking into account crucial effects such as scale-dependent bias, neutrino bias or redshift-space distortions. It can be easily re-calibrated on user-supplied simulation outputs which accuracy should match the required accuracy of the model, set, in turn,  by  the expected precision achievable from the data set of interest.  The design of the code makes it possible to use as a complement to other cosmological codes or even to add other cosmological phenomena like Alcock-Paczynski, wide-angle or GR corrections. We envision it will be useful for the analysis of next-generation surveys such as Euclid, DESI, WFIRST, SKA, PFS, EMU
and LSST.

\acknowledgments
We acknowledge support  by the Spanish MINECO under projects AYA2014-58747-P AEI/FEDER UE;  MDM-2014-0369 of ICCUB (Unidad de Excelencia Maria de Maeztu) and European Union's Horizon 2020 research and innovation programme ERC (BePreSySe, grant agreement 725327). Simulations were run at the \textbf{Rusty} cluster of the Center for Computational Astrophysics. The work of FVN is supported by the Simons Foundation.
AR has received funding from the People Programme (Marie Curie Actions) of the European Union H2020 Programme under REA grant agreement number 706896 (COSMOFLAGS). A special thank you to Cora Dvorkin, Shun Saito and Marko Simonovic for their precious insights. We acknowledge  Alexander mead and Emiliano Sefusatti for  feedback and comments on the draft. DV would like to thank his fellow PhD students (Katie, Jose, Nicolas, Ali and Sam) at ICC for their patience and support.

%%%%%%%%%%%%%%%%%% APPENDIX %%%%%%%%%%%%%%%%%%%%
%%%%%%%%%%%%%%%%%%%%%%%%%%%%%%%%%%%%%%%%%%%%
%%%%%%%%%%%%%%%%%%%%%%%%%%%%%%%%%%%%%%%%%%%%%
\appendix

\section{MCMC fitting}
\label{sec:tuning}
We used the MCMC ensemble sampler \textbf{emcee} \cite{47} to fit the  bias coefficients and  quantify their error using  least squares results as initial guesses. 

The errors on the bias  as a function of wavenumber $k$  is  given by the standard deviation of the 10 pairs of realizations.  Covariance between different $k$-bins is ignored, given the limited number of available simulations. This is justified by simplicity and  by the fact that we work in the linear and mildly-non-linear regime.
The likelihood is taken to be Gaussian.  This is a standard assumption  widely used in the literature. In reality, even if the over-density field is Gaussian (which is not because of bias and gravitational instability), its power spectrum does not follow a Gaussian probability distribution. However for band powers,  especially those populated by many modes and therefore with better signal to noise, the central limit theorem ensures that the the Gaussian approximation holds well.
The parameters to fit are the set of the bias parameters of the model, for each of the four  redshift snapshots ($n_z$) and each  of the four  mass bins ($n_M$) . Hence the total number of parameters $n_{params}$ is $n_z\times n_M\times n_{\rm model}$ where $n_{\rm model}$ is 4 for the polynomial model, 3 for the polynomial model with only even powers of $k$,  and 4 or 5 in the perturbation theory-based fits. We use uniform  improper priors for all the parameters. 

\begin{figure}[h!]
 \makebox[\textwidth]{
%\begin{center}
\includegraphics[scale=0.5]{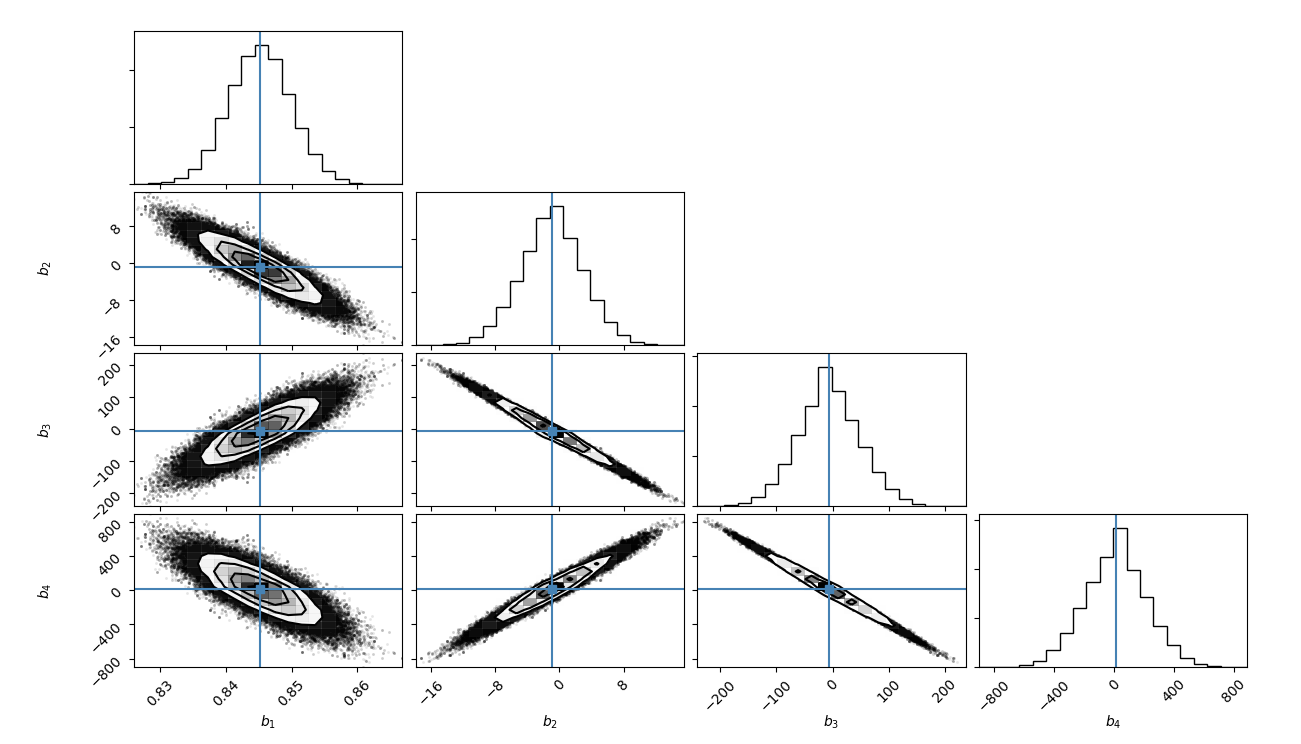}}
%\end{center}
\caption{Posteriors for  the bias coefficients for the polynomial  model, mass bin M1, z =0, $k_{\rm max}$ = 0.12 $h$/Mpc, case II}
\label{corner_pl}
\end{figure}

\begin{figure}[h!]
 \makebox[\textwidth]{
%\begin{center}
\includegraphics[scale=0.5]{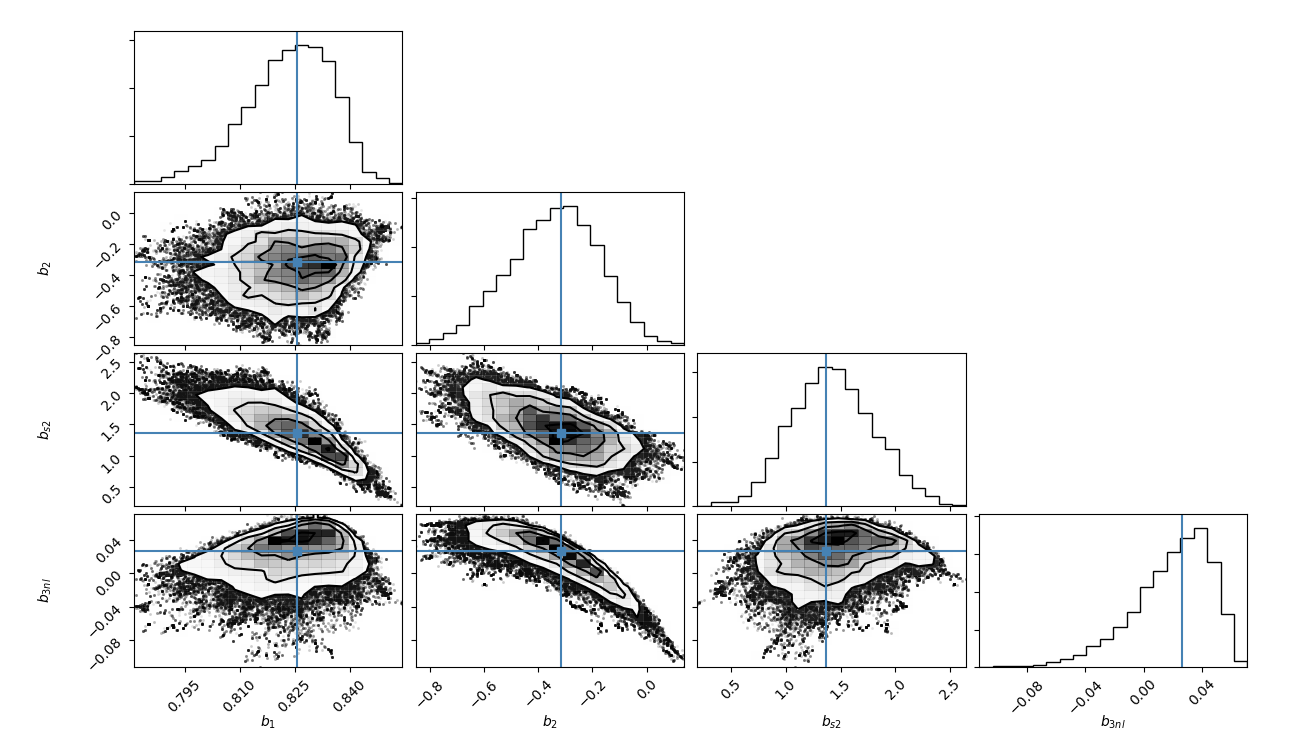}}
%\end{center}
\caption{Posteriors for the bias coefficients for the  perturbation theory-based 2nd order expansion model. Mass bin M1, z =0, kmax = 0.12 h/Mpc, case II}
\label{corner_exp}
\end{figure}

 \textbf{Emcee} used 300 walkers and 1000 steps  for each walker.
Illustrative cases for the posteriors for the bias parameters of the models considered are shown in Figs.\ref{corner_pl}-\ref{corner_exp} (see figure caption for details).

The  best  fits of each bias parameter,  and errors marginalised over all other parameters,  as function of the mass bins and redshift snapshots  are reported in tables Appendix \ref{sec:tabelsbiascoeffs1} and \ref{sec:tabelsbiascoeffs2}.
 We introduced a shot noise correction parameter,  to account for non-Poisson behaviour of shot noise,  which is marginalised over. The value of this parameter is not reported here  because it is kept as  a nuisance parameter in BE-HaPPy.

\section{Fit to the halo bias, dependence on mass bin}
\label{eachmass}
For completeness we report the ratio between the halo bias obtained from the simulations  and the fit   (Fig 2.) for each mass bin. We also report the mass deendenve of the residuals to the bias fit for (Fig 5), only for $z=1$ which is where the effect we discuss becomes evident. 
\begin{figure}[h!]
\begin{center}
\includegraphics[scale=0.497]{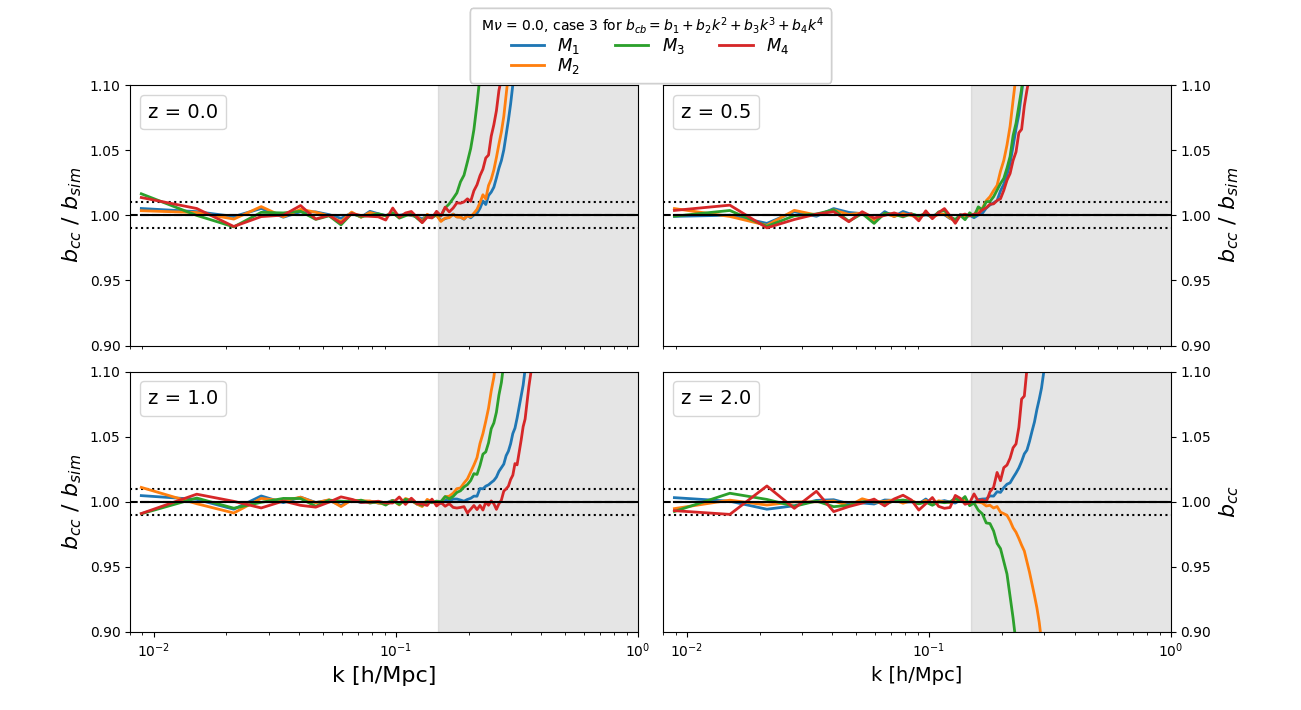}
\caption{Ratio between the halo bias obtained from the simulations and the fit split by mass bin, see Fig. 2}
\end{center}
\end{figure}
\begin{figure}[h!]
\begin{center}
\includegraphics[scale=0.3]{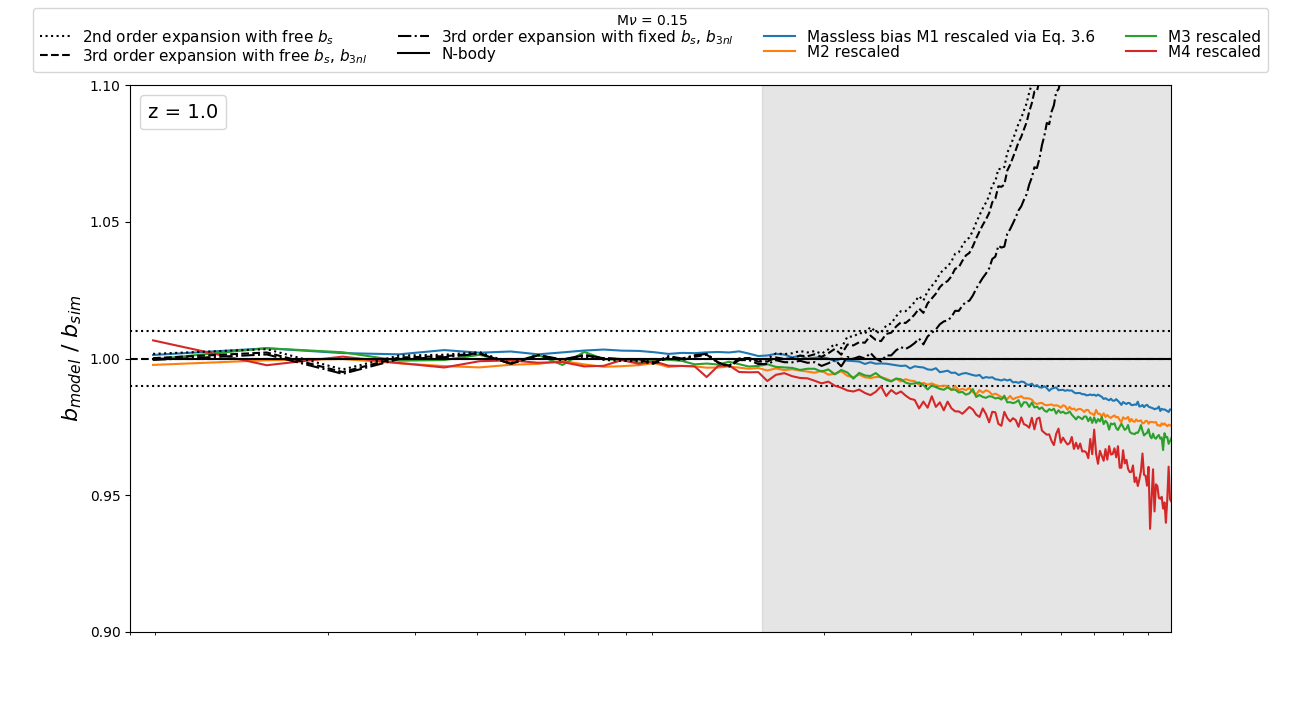}
\caption{Fig 5, bottom panel at $z=1$ but now also showing the dependence on the mass bin.}
\end{center}
\end{figure}

\section{PT terms}
\label{sec:ptterms}
For completeness we report here the expression for the perturbation theory terms used the main text. 
\begin{equation}
P_{b2,\delta}(k) = \int \frac{d^{3}q}{(2\pi)^3}P^{lin}(q)P^{lin}(|k-q|)F_{s}^{(2)}(q,k-q)
\end{equation}

\begin{equation}
P_{b2,\theta}(k) = \int \frac{d^{3}q}{(2\pi)^3}P^{lin}(q)P^{lin}(|k-q|)G_{s}^{(2)}(q,k-q)
\end{equation}

\begin{equation}
P_{bs2,\delta}(k) = \int \frac{d^{3}q}{(2\pi)^3}P^{lin}(q)P^{lin}(|k-q|)F_{s}^{(2)}(q,k-q)S	_{s}^{(2)}(q,k-q)
\end{equation}

\begin{equation}
P_{bs2,\theta}(k) = \int \frac{d^{3}q}{(2\pi)^3}P^{lin}(q)P^{lin}(|k-q|)G_{s}^{(2)}(q,k-q)S	_{s}^{(2)}(q,k-q)
\end{equation}

\begin{equation}
P_{b22}(k) = \frac{1}{2} \int \frac{d^{3}q}{(2\pi)^3} P^{lin}(q)\left[P^{lin}(|k-q|) -  P^{lin}(q)\right]
\end{equation}

\begin{equation}
P_{b2s2}(k) = -\frac{1}{2} \int \frac{d^{3}q}{(2\pi)^3} P^{lin}(q)\left[\frac{2}{3} P^{lin}(q) - P^{lin}(|k-q|)S_{s}^{(2)}(q,k-q)\right]
\end{equation}

\begin{equation}
P_{bs22}(k) = -\frac{1}{2} \int \frac{d^{3}q}{(2\pi)^3} P^{lin}(q)\left[\frac{4}{9} P^{lin}(q) - P^{lin}(|k-q|)S_{s}^{(2)}(q,k-q)^2\right]
\end{equation}

where $F_{s}^{(2)}$, $G_{s}^{(2)}$ and $S_{s}^{(2)}$ are 2nd order Perturbation Theory kernels.

\section{Bias coefficients: polynomial fit}
\label{sec:tabelsbiascoeffs1}
In Tab.\ref{coeffplod1},\ref{coeffplod2},\ref{coeffplod3} and \ref{coeffplod4} we report the best fit bias coefficients and their marginal errors for the polynomial model of Sec.~\ref{sec:scale-dependent-bias}.

\begin{table}[h!]
\begin{center}
\resizebox{\columnwidth}{!}{
\begin{tabular}{|c|c|c|c|c|c|c|c|c|c|c|c|c|}
\hline
Mass bins & $b_1$ & --err $b_1$  &  +err $b_1$  & $b_2$ & -- err $b_2$ & + err $b_2$  & $b_3$ & -- err $b_3$ & + err $b_3$ & $b_4$ & --err $b_4$ & + err $b_4$  \\ \hline
M1 & \textbf{0.845} & 0.005 & 0.005 & \textbf{-0.898} & 3.833 & 3.81 & \textbf{-6.75} & 53.83 & 54.688 & \textbf{12.125} & 205.818 & 203.14 \\
M2 & \textbf{0.888} & 0.006 & 0.006 & \textbf{0.893} & 4.916 & 4.924 & \textbf{-21.222} & 69.02 & 68.804 & \textbf{42.956} & 254.91 & 256.332 \\
M3 & \textbf{1.023} & 0.008 & 0.008 & \textbf{5.545} & 6.222 & 6.225 & \textbf{-79.71} & 86.735 & 87.672 & \textbf{245.958} & 326.821 & 322.981 \\
M4 & \textbf{1.29} & 0.012 & 0.012 & \textbf{1.655} & 10.178 & 10.163 & \textbf{-38.373} & 145.538 & 144.508 & \textbf{102.042} & 545.552 & 555.432 \\ \hline
\end{tabular}}
\end{center}
\caption{Polynomial model coefficients of $b_{\rm cc}$; $k_{\rm max}$ = 0.15 h/Mpc, z = 0.0.}
\label{coeffplod1}
\end{table}

\begin{table}[h!]
\begin{center}
\resizebox{\columnwidth}{!}{
\begin{tabular}{|c|c|c|c|c|c|c|c|c|c|c|c|c|}
\hline
Mass bins & $b_1$ & --err $b_1$  &  +err $b_1$  & $b_2$ & -- err $b_2$ & + err $b_2$  & $b_3$ & -- err $b_3$ & + err $b_3$ & $b_4$ & --err $b_4$ & + err $b_4$  \\ \hline
M1 & \textbf{1.04} & 0.006 & 0.006 & \textbf{2.919} & 4.711 & 4.682 & \textbf{-38.52} & 65.679 & 67.014 & \textbf{108.898} & 253.156 & 245.006 \\
M2 & \textbf{1.127} & 0.008 & 0.008 & \textbf{6.157} & 6.454 & 6.554 & \textbf{-77.288} & 91.978 & 91.83 & \textbf{242.857} & 347.767 & 346.377 \\
M3 & \textbf{1.366} & 0.009 & 0.009 & \textbf{6.914} & 7.61 & 7.686 & \textbf{-86.903} & 108.453 & 109.369 & \textbf{279.133} & 415.68 & 410.437 \\
M4 & \textbf{1.792} & 0.013 & 0.013 & \textbf{1.792} & 10.659 & 10.54 & \textbf{-32.749} & 155.715 & 156.452 & \textbf{106.696} & 612.877 & 616.677 \\ \hline
\end{tabular}}
\end{center}
\caption{Polynomial model coefficients of $b_{\rm cc}$ ; $k_{\rm max}$ = 0.15 h/Mpc, z = 0.5.}
\label{coeffplod2}
\end{table}

\begin{table}[h!]
\begin{center}
\resizebox{\columnwidth}{!}{
\begin{tabular}{|c|c|c|c|c|c|c|c|c|c|c|c|c|}
\hline
Mass bins & $b_1$ & --err $b_1$  &  +err $b_1$  & $b_2$ & -- err $b_2$ & + err $b_2$  & $b_3$ & -- err $b_3$ & + err $b_3$ & $b_4$ & --err $b_4$ & + err $b_4$  \\ \hline
M1 & \textbf{1.332} & 0.007 & 0.007 & \textbf{2.15} & 5.736 & 5.676 & \textbf{-15.717} & 78.401 & 79.434 & \textbf{24.882} & 293.963 & 287.496 \\
M1 & \textbf{1.487} & 0.008 & 0.008 & \textbf{3.348} & 6.889 & 6.695 & \textbf{-34.19} & 93.048 & 97.687 & \textbf{107.953} & 367.823 & 346.865 \\
M1 & \textbf{1.868} & 0.012 & 0.012 & \textbf{1.282} & 9.469 & 9.532 & \textbf{-1.931} & 136.007 & 134.97 & \textbf{-13.002} & 512.986 & 517.966 \\
M1 & \textbf{2.503} & 0.019 & 0.019 & \textbf{8.383} & 17.182 & 17.466 & \textbf{-61.065} & 260.602 & 254.053 & \textbf{131.255} & 990.305 & 1022.662 \\ \hline
\end{tabular}}
\end{center}
\caption{Polynomial model coefficients of $b_{\rm cc}$; $k_{\rm max}$ = 0.15 h/Mpc, z = 1.0.}
\label{coeffplod3}
\end{table}

\begin{table}[h!]
\begin{center}
\resizebox{\columnwidth}{!}{
\begin{tabular}{|c|c|c|c|c|c|c|c|c|c|c|c|c|}
\hline
Mass bins & $b_1$ & --err $b_1$  &  +err $b_1$  & $b_2$ & -- err $b_2$ & + err $b_2$  & $b_3$ & -- err $b_3$ & + err $b_3$ & $b_4$ & --err $b_4$ & + err $b_4$  \\ \hline
M1 & \textbf{2.199} & 0.009 & 0.009 & \textbf{3.825} & 7.197 & 7.158 & \textbf{-21.639} & 101.805 & 102.2 & \textbf{77.61} & 388.068 & 387.347 \\
M2 & \textbf{2.531} & 0.013 & 0.012 & \textbf{2.094} & 9.894 & 9.993 & \textbf{25.194} & 141.411 & 141.223 & \textbf{-113.229} & 539.035 & 535.428 \\
M3 & \textbf{3.297} & 0.029 & 0.029 & \textbf{7.211} & 22.679 & 22.397 & \textbf{67.113} & 317.97 & 324.107 & \textbf{-398.872} & 1244.717 & 1206.446 \\
M4 & \textbf{4.664} & 0.086 & 0.085 & \textbf{45.85} & 70.202 & 70.9 & \textbf{-147.646} & 1025.801 & 1017.184 & \textbf{96.813} & 3901.527 & 3973.001 \\ \hline
\end{tabular}}
\end{center}
\caption{ Polynomial model  coefficients of $b_{\rm cc}$; $k_{\rm max}$ = 0.15 h/Mpc, z = 2.0.}
\label{coeffplod4}
\end{table}

\section{Third-order bias}
\label{appendix:b3}
We compute the coefficient of the 3rd-order non local bias term
using Eq. 53 of \cite{37}.  As introduced and explained in McDonald \& Roy \cite{40}, in the expansion of the power spectrum, the three integrals involving the third-order nonlocal terms are exactly proportional to each other after renormalization,  and can be  encompassed in a single third order bias term $b_{3nl}$, simplifying  significantly the resulting expressions.  Thus we  just need to compute the quantity $\sigma_{3}^{2}$:

\begin{equation}
\sigma_{3}^{2} =  \frac{105}{16}\int \frac{d^{3}q}{(2\pi)^{3}} P^{lin}(q)\left[D^{(2)}(-q,k)S^{(2)}(q,k-q)+\frac{8}{63}\right]
\label{3nl}
\end{equation}
Through a   change of variable we can rewrite the expression $\sigma_{3}^{2} \times P^{lin}(k)$ as
\begin{equation}
\begin{split}
\sigma_{3}^{2}P^{lin}(k) &=  \frac{105}{16}P^{lin}(k)\int \frac{d^{3}q}{(2\pi)^{3}} P^{lin}(q)\left[D^{(2)}(-q,k)S^{(2)}(q,k-q)+\frac{8}{63}\right]\\
&=  \frac{105\:k^{3}}{16\:(2\pi)^{2}}P^{lin}(k)\int dr\:r^{2} P^{lin}(kr)I_{R}(r)
\end{split}
\label{3nlbis}
\end{equation}

where \[I_{R} = \int_{-1}^{1} \left[D^{(2)}(-q,k)S^{(2)}(q,k-q)+\frac{8}{63}\right]d\mu\] 
and $r = {q}/{k}$; $\mu = {\vec{k}\:.\:\vec{q}}/(kq)$.

The second line of Eq. (\ref{3nlbis})  is very  similar to a $P_{13}$ convolution integral (see section 2.3 of \cite{8}) simplifying the  implementation in FAST-PT.

\begin{figure}[h!]
\begin{center}
\includegraphics[scale=0.55]{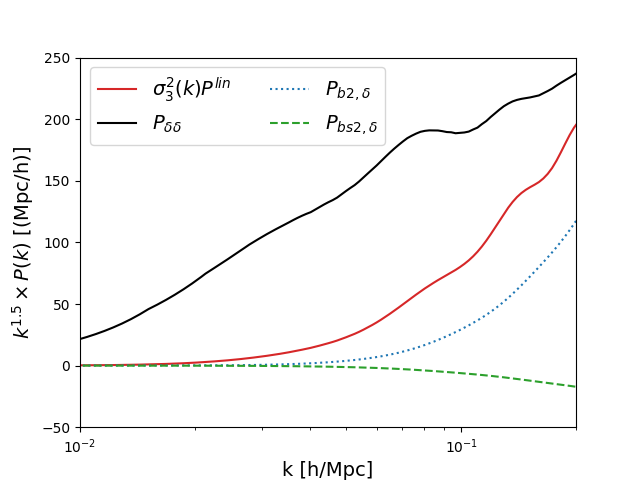} 
\end{center}
\caption{Comparison of terms  in the 3rd order expansion. This figure can be compared directly with Fig. 1 of Ref. \cite{38}, bearing in mind that here the power spectra are in units of Mpc/h.}
\label{saito}
\end{figure}

The relevant terms appearing in  Eq. \ref{bpt1}  --$\sigma_{3}^{2}P^{lin}(k)$,  the non linear matter power spectrum from simulation $P_{\delta\delta}$, the second-order local bias term $P_{b2,\delta}$ and the second-order nonlocal bias term $P_{bs2,\delta}$-- are shown in Fig.~\ref{saito}. Like Ref.~\cite{39} we see that  the third-order nonlocal term dominate over the second-order local and nonlocal terms, as long as  the $b_{2}$ term is sufficiently small.

\section{Bias coefficients: perturbation theory-based fit}
\label{sec:tabelsbiascoeffs2}
Here we report the  the best fit bias coefficients and their marginal errors for the perturbation theory-based  model of Sec.~\ref{sec:PTbias}.

\begin{table}[h!]
\begin{center}
\resizebox{\columnwidth}{!}{
\begin{tabular}{|c|c|c|c|c|c|c|c|c|c|}
\hline

Mass bins & $b_1$ & --err $b_1$  &  +err $b_1$  & $b_2$ & -- err $b_2$ & + err $b_2$  & $b_{s2}$ & -- err $b_{s2}$ & + err $b_{s2}$  \\ \hline
M1 & \textbf{0.838} & 0.006 & 0.008 & \textbf{-0.328} & 0.086 & 0.095 & \textbf{0.038} & 0.143 & 0.08 \\
M2 & \textbf{0.849} & 0.019 & 0.018 & \textbf{-0.227} & 0.084 & 0.158 & \textbf{1.696} & 0.481 & 0.436 \\
M3 & \textbf{0.98} & 0.024 & 0.024 & \textbf{-0.198} & 0.082 & 0.141 & \textbf{1.907} & 0.531 & 0.493 \\
M4 & \textbf{1.277} & 0.017 & 0.02 & \textbf{-0.474} & 0.196 & 0.253 & \textbf{-0.175} & 0.295 & 0.171 \\ \hline
\end{tabular}}
\end{center}
\caption{Coefficients of $b_{\rm cc}$,  2nd order expansion model, $k_{\rm max}$ = 0.15 h/Mpc, z= 0.}
\label{coeff2a}
\end{table}

\begin{table}[h!]
\begin{center}
\resizebox{\columnwidth}{!}{
\begin{tabular}{|c|c|c|c|c|c|c|c|c|c|}
\hline
Mass bins & $b_1$ & --err $b_1$  &  +err $b_1$  & $b_2$ & -- err $b_2$ & + err $b_2$  & $b_{s2}$ & -- err $b_{s2}$ & + err $b_{s2}$  \\ \hline
M1 &\textbf{1.032} & 0.01 & 0.012 & \textbf{-0.362} & 0.152 & 0.19 & \textbf{-0.374} & 0.221 & 0.124 \\
M2 &\textbf{1.118} & 0.015 & 0.016 & \textbf{-0.37} & 0.22 & 0.279 & \textbf{-0.577} & 0.297 & 0.153 \\
M3 &\textbf{1.344} & 0.024 & 0.03 & \textbf{-0.044} & 0.083 & 0.171 & \textbf{1.948} & 0.992 & 1.058 \\
M4 &\textbf{1.79} & 0.01 & 0.022 & \textbf{-0.302} & 0.235 & 0.439 & \textbf{-0.286} & 0.884 & 0.44 \\ \hline
\end{tabular}}
\end{center}
\caption{Coefficients of $b_{\rm cc}$, second order expansion model, $k_{\rm max}$ = 0.15 h/Mpc, z= 0.5}
\label{coeff2b}
\end{table}

\begin{table}[h!]
\begin{center}
\resizebox{\columnwidth}{!}{
\begin{tabular}{|c|c|c|c|c|c|c|c|c|c|}
\hline
Mass bins & $b_1$ & --err $b_1$  &  +err $b_1$  & $b_2$ & -- err $b_2$ & + err $b_2$  & $b_{s2}$ & -- err $b_{s2}$ & + err $b_{s2}$  \\ \hline
M1 &\textbf{1.321} & 0.012 & 0.019 & \textbf{0.093} & 0.058 & 0.096 & \textbf{1.455} & 0.966 & 1.012 \\
M2 &\textbf{1.488} & 0.005 & 0.012 & \textbf{0.001} & 0.158 & 0.27 & \textbf{-0.753} & 0.837 & 0.483 \\
M3 &\textbf{1.869} & 0.007 & 0.014 & \textbf{0.094} & 0.164 & 0.317 & \textbf{-0.696} & 1.39 & 0.808 \\
M4 &\textbf{2.481} & 0.03 & 0.041 & \textbf{-0.231} & 0.653 & 0.968 & \textbf{-2.719} & 1.334 & 0.61 \\ \hline
\end{tabular}}
\end{center}
\caption{Coefficients of $b_{\rm cc}$, second order expansion model,  $k_{\rm max}$ = 0.15 h/Mpc, z= 1.0.}
\label{coeff2c}
\end{table}

\begin{table}[h!]
\begin{center}
\resizebox{\columnwidth}{!}{
\begin{tabular}{|c|c|c|c|c|c|c|c|c|c|}
\hline
Mass bins & $b_1$ & --err $b_1$  &  +err $b_1$  & $b_2$ & -- err $b_2$ & + err $b_2$  & $b_{s2}$ & -- err $b_{s2}$ & + err $b_{s2}$  \\ \hline
M1 &\textbf{2.201} & 0.005 & 0.011 & \textbf{0.983} & 0.137 & 0.283 & \textbf{-1.45} & 2.049 & 1.674 \\
M2 &\textbf{2.532} & 0.007 & 0.013 & \textbf{1.4} & 0.204 & 0.424 & \textbf{-2.437} & 2.557 & 1.91 \\
M3 &\textbf{3.264} & 0.029 & 0.067 & \textbf{1.995} & 0.794 & 1.364 & \textbf{-9.241} & 3.797 & 1.872 \\
M4 &\textbf{4.611} & 0.093 & 0.236 & \textbf{7.502} & 0.759 & 1.452 & \textbf{-1.539} & 12.318 & 10.691 \\ \hline
\end{tabular}}
\end{center}
\caption{Coefficients of $b_{\rm cc}$,  second order expansion model, $k_{\rm max}$ = 0.15 h/Mpc, z= 2.0.}
\label{coeff2d}
\end{table}

\begin{table}[h!]
\begin{center}
\resizebox{\columnwidth}{!}{
\begin{tabular}{|c|c|c|c|c|c|c|c|c|c|c|c|c|}
\hline

Mass bins & $b_1$ & --err $b_1$  &  +err $b_1$  & $b_2$ & -- err $b_2$ & + err $b_2$  & $b_{s2}$ & -- err $b_{s2}$ & + err $b_{s2}$ & $b_{3nl}$ & -- err $b_{3nl}$ & + err $b_{3nl}$ \\ \hline
M1 &\textbf{0.826} & 0.009 & 0.013 & \textbf{-0.315} & 0.158 & 0.183 & \textbf{1.368} & 0.352 & 0.353 & \textbf{0.026} & 0.02 & 0.03 \\
M2 &\textbf{0.855} & 0.015 & 0.016 & \textbf{-0.444} & 0.222 & 0.206 & \textbf{1.879} & 0.462 & 0.403 & \textbf{0.041} & 0.026 & 0.033 \\
M3 &\textbf{0.989} & 0.017 & 0.022 & \textbf{-0.552} & 0.26 & 0.273 & \textbf{2.211} & 0.541 & 0.535 & \textbf{0.068} & 0.028 & 0.037 \\
M4 &\textbf{1.266} & 0.015 & 0.025 & \textbf{-0.698} & 0.368 & 0.361 & \textbf{2.211} & 0.759 & 0.78 & \textbf{0.093} & 0.027 & 0.047 \\ \hline
\end{tabular}}
\end{center}
\caption{Coefficients of $b_{\rm cc}$, third expansion model with $b_{3nl}$ kept as free parameter, $k_{max}$ = 0.15 h/Mpc, z = 0.}
\label{coeff3a}
\end{table}

\begin{table}[h!]
\begin{center}
\resizebox{\columnwidth}{!}{
\begin{tabular}{|c|c|c|c|c|c|c|c|c|c|c|c|c|}
\hline
Mass bins & $b_1$ & --err $b_1$  &  +err $b_1$  & $b_2$ & -- err $b_2$ & + err $b_2$  & $b_{s2}$ & -- err $b_{s2}$ & + err $b_{s2}$ & $b_{3nl}$ & -- err $b_{3nl}$ & + err $b_{3nl}$ \\ \hline
M1 & \textbf{1.029} & 0.007 & 0.011 & \textbf{-0.562} & 0.284 & 0.293 & \textbf{1.848} & 0.556 & 0.585 & \textbf{0.097} & 0.023 & 0.041 \\
M2 &\textbf{1.118} & 0.007 & 0.014 & \textbf{-0.925} & 0.407 & 0.396 & \textbf{2.339} & 0.851 & 0.773 & \textbf{0.136} & 0.03 & 0.051 \\
M3 &\textbf{1.358} & 0.008 & 0.014 & \textbf{-0.634} & 0.388 & 0.439 & \textbf{1.907} & 1.011 & 1.011 & \textbf{0.155} & 0.035 & 0.067 \\
M4 &\textbf{1.788} & 0.012 & 0.012 & \textbf{-0.466} & 0.544 & 0.59 & \textbf{1.061} & 1.337 & 1.345 & \textbf{0.094} & 0.074 & 0.138 \\ \hline
\end{tabular}}
\end{center}
\caption{Coefficients of $b_{\rm cc}$, third  order expansion model with $b_{3nl}$ kept as free parameter, $k_{max}$ = 0.15 h/Mpc, z = 0.5.}
\label{coeff3b}
\end{table}

\begin{table}[h!]
\begin{center}
\resizebox{\columnwidth}{!}{
\begin{tabular}{|c|c|c|c|c|c|c|c|c|c|c|c|c|}
\hline
Mass bins & $b_1$ & --err $b_1$  &  +err $b_1$  & $b_2$ & -- err $b_2$ & + err $b_2$  & $b_{s2}$ & -- err $b_{s2}$ & + err $b_{s2}$ & $b_{3nl}$ & -- err $b_{3nl}$ & + err $b_{3nl}$ \\ \hline
M1 &\textbf{1.329} & 0.006 & 0.006 & \textbf{-0.354} & 0.382 & 0.417 & \textbf{0.799} & 1.008 & 0.927 & \textbf{0.136} & 0.06 & 0.11 \\
M2 &\textbf{1.488} & 0.008 & 0.008 & \textbf{-0.116} & 0.397 & 0.442 & \textbf{0.218} & 1.105 & 1.141 & \textbf{0.087} & 0.116 & 0.152 \\
M3 &\textbf{1.866} & 0.011 & 0.011 & \textbf{-0.091} & 0.56 & 0.591 & \textbf{0.367} & 1.588 & 1.591 & \textbf{0.105} & 0.169 & 0.21 \\
M4 & \textbf{2.5} & 0.017 & 0.017 & \textbf{-0.551} & 1.146 & 1.286 & \textbf{1.261} & 2.919 & 2.685 & \textbf{0.358} & 0.167 & 0.323 \\ \hline
\end{tabular}}
\end{center}
\caption{Coefficients of $b_{\rm cc}$,  third  order expansion model with $b_{3nl}$ kept as free parameter, $k_{max}$ = 0.15 h/Mpc, z = 1.0.}
\label{coeff3c}
\end{table}

\begin{table}[h!]
\begin{center}
\resizebox{\columnwidth}{!}{
\begin{tabular}{|c|c|c|c|c|c|c|c|c|c|c|c|c|}
\hline
Mass bins & $b_1$ & --err $b_1$  &  +err $b_1$  & $b_2$ & -- err $b_2$ & + err $b_2$  & $b_{s2}$ & -- err $b_{s2}$ & + err $b_{s2}$ & $b_{3nl}$ & -- err $b_{3nl}$ & + err $b_{3nl}$ \\ \hline
M1 &\textbf{2.202} & 0.009 & 0.009 & \textbf{1.085} & 0.738 & 0.728 & \textbf{-1.57} & 1.898 & 1.818 & \textbf{-0.054} & 0.338 & 0.419 \\
M2 &\textbf{2.533} & 0.013 & 0.013 & \textbf{1.516} & 0.863 & 0.984 & \textbf{-2.665} & 2.566 & 2.254 & \textbf{-0.074} & 0.513 & 0.556 \\
M3 &\textbf{3.273} & 0.036 & 0.05 & \textbf{3.175} & 1.781 & 2.094 & \textbf{0.042} & 4.611 & 4.333 & \textbf{-0.006} & 1.158 & 1.347 \\
M4 &\textbf{4.624} & 0.085 & 0.123 & \textbf{6.026} & 4.376 & 5.034 & \textbf{-1.637} & 11.736 & 9.845 & \textbf{1.528} & 2.591 & 3.921 \\ \hline
\end{tabular}}
\end{center}
\caption{Coefficient of $b_{\rm cc}$,  third  order expansion model with $b_{3nl}$ kept as free parameter, $k_{\rm max}$ = 0.15 h/Mpc, z = 2.0.}
\label{coeff3d}
\end{table}

\section{redshift-space checks}
\label{sec:redshift-space}
We have performed a cross check of our implementation of redshift-space distortions in FAST-PT with the original results by  Scoccimarro   et al.  \cite{42}.

\begin{figure}[h!]
 \makebox[\textwidth]{
\includegraphics[scale=0.45]{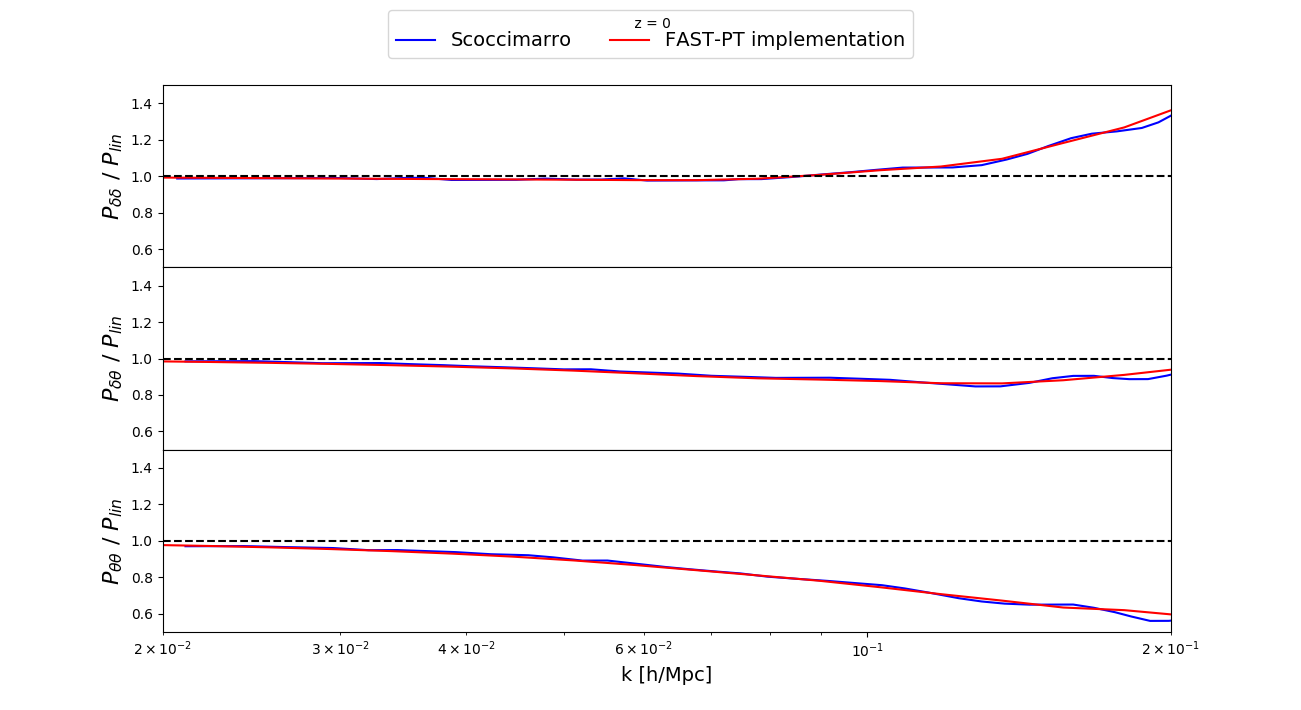}}
\caption{Comparison of FAST-PT modifications (red) vs Scoccimarro   \cite{42} (blue) at z = 0.  For this figure we use a flat $\Lambda$CDM model with $\Omega_{m}$ = 0.26, $\sigma_{8}$ = 0.9, $\Omega_{b}$ = 0.04 and $h$ = 0.7 as in \cite{42}}
\label{scocci}
\end{figure} 

Figure~\ref{scocci} shows  excellent agreement with only a  little  residual discrepancy at small scales, which is probably due to the fact that we do not know exactly  (and therefore may not have matched perfectly) all the   cosmological parameters used in \cite{42} to initialize the input linear power spectrum.

\section{Multipole expansion}

redshift-space power spectra are often plotted in terms of their  so-called multipoles.
Both Eqs. \ref{sco} and \ref{etns} depend on $k$ and $\mu$. Instead of working with these 2-dimensional functions, we expand them into Legendre polynomials, $L_\ell$, following the traditional approach:

\begin{equation}
P_{l}^{s}(k) = \frac{2l+1}{2} \int_{-1}^{1} d\mu \: P^{s}(k,\mu)L_{l}(\mu)\\
\end{equation}
In order to isolate the $\mu$ dependence of the $A$ and $B$ coefficients of the TNS and eTNS models, we write

\[A(k,\mu,f) = \bar{A}(k,\mu, f) \times k\mu f\] \[B(k,\mu,f) = \bar{B}(k,\mu, f) \times (k\mu f)^2\]
The $\bar{A}$ and $\bar{B}$ integrals can be decomposed as a summation of convolution integrals (see Appendix C of \cite{8})
which in turn can be written as Legendre expansions.
\[\bar{A}(k,\mu,f) = \sum_{i=0}A_{i}(k,f)\mu^i\]
\[\bar{B}(k,\mu,f) = \sum_{i=0}B_{i}(k,f)\mu^i\]

We can this finally express $A$ and $B$ as 
\[A(k,\mu,f) = k f \times \sum_{i=0}A_{i}(k,f)\mu^{i+1}\] 
\[B(k,\mu,f) = (k f)^2 \times \sum_{i=0}B_{i}(k,f)\mu^{i+2}.\]

For biased tracers in the above equations $f\longrightarrow \beta$.
In \textbf{BE-HaPPy}  the integration of $\mu$  is split in the same way as Cole et al. \cite{46}:

\begin{equation}
\begin{split}
P_{l}^{s}(k) &= \frac{2l+1}{2} \int_{-1}^{1} d\mu \: P_{g}^{s}(k,\mu)L_{l}(\mu)\\
&= \frac{2l+1}{2} \int_{-1}^{1} d\mu \: K(k,\mu)F(k,\mu^{2})L_{l}(\mu)~,
\end{split}
\end{equation}
where $L_{l}(\mu)$ are the first even Legendre polynomials, and $K(k,\mu)$ can be the Kaiser, Scoccimarro or TNS models, and $F(k,\mu^{2})$ is the FoG term. \textbf{BE-HaPPy} allows the user to choose among all these different models.

\section{$\sigma_8$ scale-dependence}
\label{sec:sigma8}
In Figure \ref{tinker3} we rescaled the amplitude of the bias calibrated with a massless neutrinos simulation and compare with the bias of a massive neutrinos simulation ($M_{\nu}=$0.15 eV). If the bias is calibrated with respect to CDM + baryons $b_{\rm cc}$, we expect that all the scale dependence is encompassed in the massless case and that the only effect of massive neutrinos would be on the amplitude of the bias. However in Figure \ref{tinker3}, we observe an extra scale dependence. This scale dependence becomes more pronounced with increasing linear bias.
Because of the degeneracy between  $M_\nu$ and $\sigma_8$, we argue that this scale dependence is not due neutrinos but to a difference of $\sigma_8$ between the massless and massive simulation. To test this we compared our rescaling procedure with another simulation (with massless neutrinos) where $\sigma_8$ is closer to  that of  the massive neutrinos simulation. All other parameters ($\Omega_m$, $\Omega_b$, $\Omega_\Lambda$, $n_s$, $h$) are identical. This is shown in Figure \ref{sig8}. One can appreciate that 
 the extra scale dependence decreases for a better matched  $\sigma_8$ (blue line). The  blue line  is only plotted to larger scale (smaller $k$)  because of the different  (lower) resolution of the simulation. Unfortunately we are  limited by the simulations available  at this time and this could not be investigated further.
\begin{figure*}
\begin{center}
\includegraphics[scale=0.47]{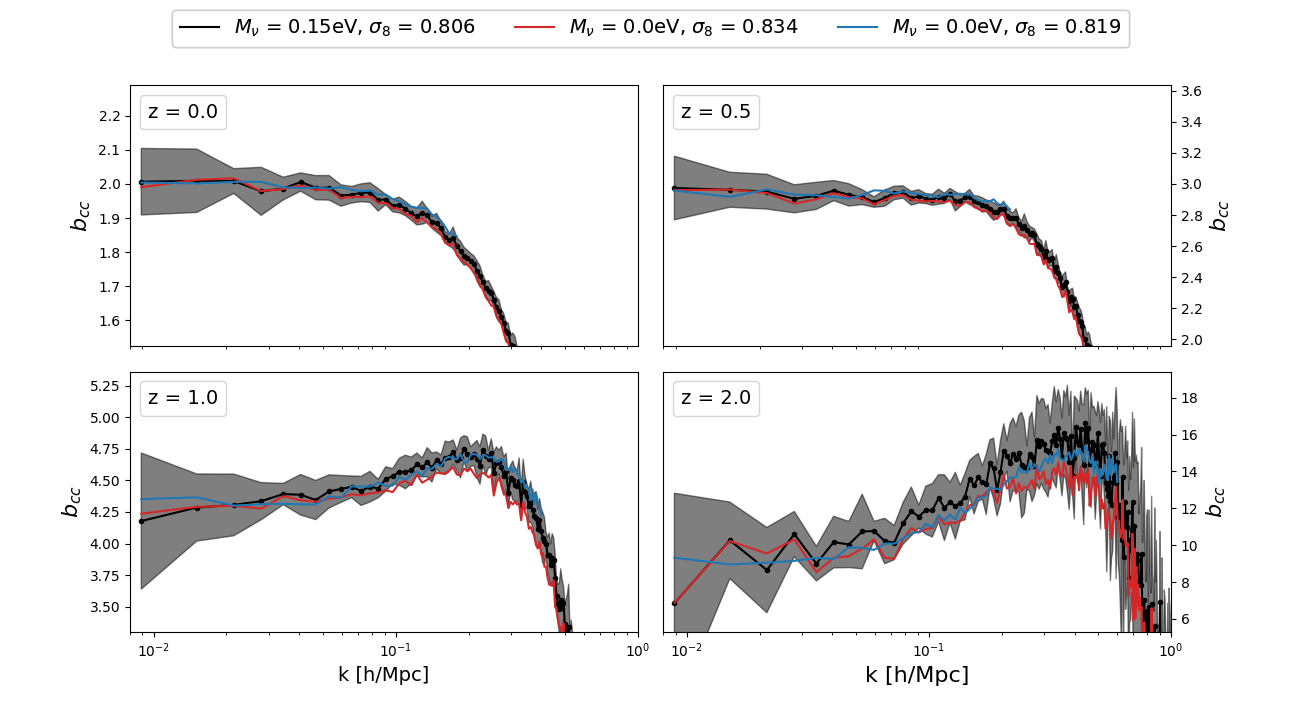}
\caption{Influence of $\sigma_8$ in the performance of  the rescaling  of equation \ref{rule3}. To enhance the effect,  the figure  corresponds to haloes with masses $> 5 \times 10^{13} M_{\odot}$.  We applied the rescaling method of Eq.~\ref{rule3} to simulations with the same cosmological parameters $\Omega_{\rm m}=\Omega_{\rm c}+\Omega_{\rm b}+\Omega_\nu=0.3175$, $\Omega_{\rm b}=0.049$, $\Omega_\Lambda=0.6825$, $\Omega_{\rm k}=0$, $h=0.6711$, $n_s=0.9624$ and $M_\nu$ = 0.0 eV but with different $\sigma_8$ (0.819 for the blue curve and 0.834 for the red one). The improvement at small scales for the lower $\sigma_8$ confirms our hypothesis about the scale dependence seen in Fig.~\ref{tinker3}.}
\label{sig8}
\end{center}
\end{figure*}

\section{Rescaling coefficient $\alpha$}
For completeness here we show the  dependence of $\alpha$ values (Eq.~\ref{analytic_bias}) on neutrino mass and redshift.
\label{sec:alpha}
\begin{figure*}
\begin{center}
\includegraphics[scale=0.47]{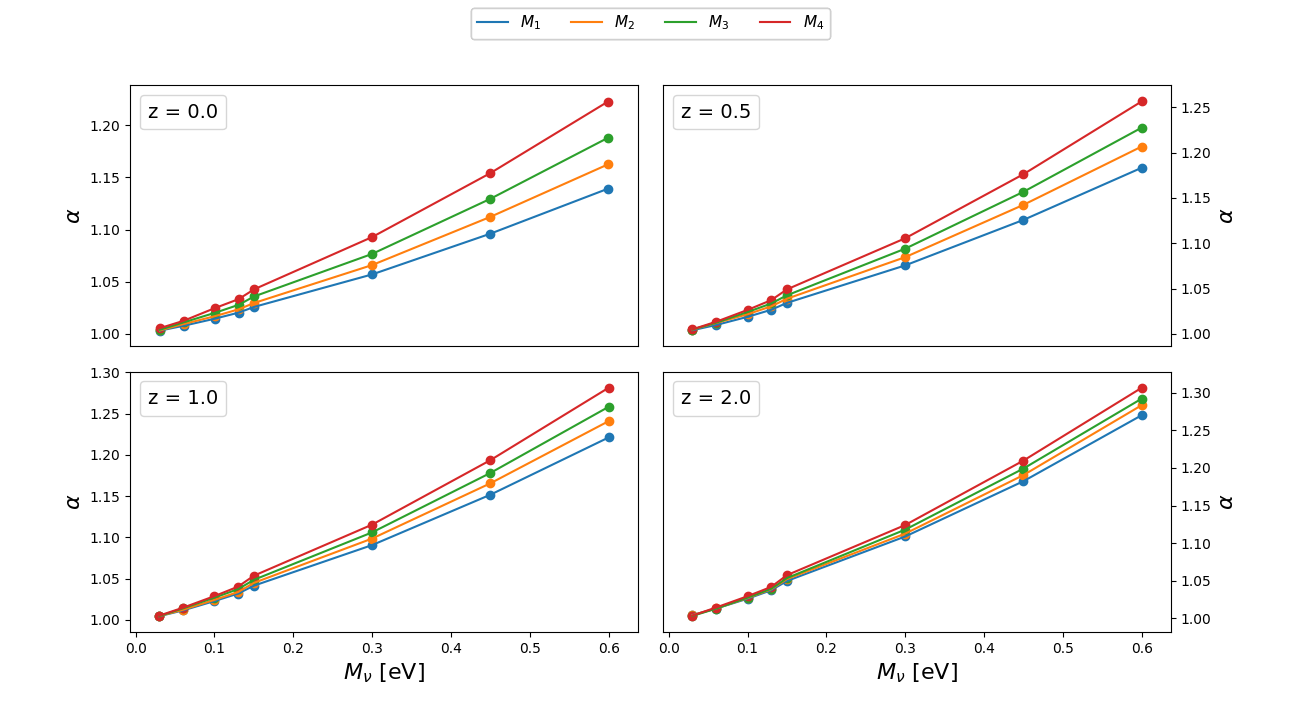}
\caption{Value of the rescaling coefficients $\alpha$ as a function of mass bins and redshift. We can see here that the relation between the rescaling and the mass of the neutrinos is quasi linear.}
\label{sig8}
\end{center}
\end{figure*}
%%%%%%%%%%%%%%%%%%%%%%%%%%%%%%%%%%%%%

\bibliography{be_happy} 
\bibliographystyle{utcaps}

\end{document}